\documentclass[
    prl,
    twocolumn,
    superscriptaddress,
    nofootinbib,
    amsmath,
    amssymb,
    aps,
    floatfix,
    balancelastpage
]{revtex4-2}

\usepackage[dvipsnames]{xcolor}
\definecolor{CiteBlue}{RGB}{45,52,151}

\usepackage[
    colorlinks=true,
    linkcolor=CiteBlue,
    urlcolor=CiteBlue,
    citecolor=CiteBlue
]{hyperref}

\usepackage[utf8]{inputenc}
\usepackage[T1]{fontenc}

\usepackage{natbib,bm,graphicx,multirow,setspace}
\usepackage[capitalise]{cleveref}

\usepackage{siunitx}
\DeclareSIUnit{\year}{yr}

\usepackage{mathrsfs}

\newcommand{\refcite}[1]{Ref.~\cite{#1}}
\newcommand{\refscite}[1]{Refs.~\cite{#1}}

\newcommand{\du}{\mathrm{d}}
\newcommand{\dd}{\,\du}
\newcommand{\bb}[1]{\bm{\mathrm{#1}}}
\newcommand{\del}{\partial}
\renewcommand{\Re}{\operatorname{Re}}
\renewcommand{\Im}{\operatorname{Im}}

\usepackage{relsize}

\newcommand{\dm}{\chi}
\newcommand{\med}{\phi}
\newcommand{\fermi}{\mathrm{F}}

\newcommand{\rpa}{^{(\mathrm{RPA})}}
\newcommand{\lindhard}{\mathrm{L}}
\newcommand{\qp}{\mathrm{QP}}
\newcommand{\ph}{\mathrm{ph}}
\newcommand{\BCS}{\mathrm{BCS}}
\newcommand{\FBCS}{{\mathcal F_\BCS}}
\newcommand{\mstar}{m_*}

\newcommand{\qpa}{{{\rm 1}}}
\newcommand{\qpb}{{{\rm 2}}}
\newcommand{\qpi}{{{\rm i}}}
\newcommand{\costh}{{\cos\theta}}
\newcommand{\Erelative}{{\mathcal{E}}}
\newcommand{\omegaph}{\omega}

\def\myhbar{}
\def\hbarsq{}
\def\csq{}
\def\overall{}

\begin{document}

\title{Directional Detection of Light Dark Matter in Superconductors}

\author{Yonit Hochberg}
\email{yonit.hochberg@mail.huji.ac.il}

\author{Eric David Kramer}
\email{ericdavidkramer@gmail.com}
\affiliation{Racah Institute of Physics, Hebrew University of Jerusalem, Jerusalem 91904, Israel}

\author{Noah Kurinsky}
\email{kurinsky@slac.stanford.edu}
\affiliation{Fermi National Accelerator Laboratory, Batavia, IL 60510, USA}
\affiliation{Kavli Institute for Cosmological Physics, University of Chicago, Chicago, IL 60637, USA}
\affiliation{SLAC National Accelerator Laboratory, Menlo Park, CA 94025}

\author{Benjamin V. Lehmann}
\email{benvlehmann@gmail.com}
\affiliation{Department of Physics, 1156 High St., University of California Santa Cruz, Santa Cruz, CA 95064, USA}
\affiliation{Santa Cruz Institute for Particle Physics, 1156 High St., Santa Cruz, CA 95064, USA}

\begin{abstract}\ignorespaces
    Superconducting detectors have been proposed as outstanding targets for the direct detection of light dark matter scattering at masses as low as a keV. We study the prospects for directional detection of dark matter in isotropic superconducting targets from the angular distribution of excitations produced in the material. We find that dark matter scattering produces initial excitations with an anisotropic distribution, and further show that this directional information can be preserved as the initial excitations relax. Our results demonstrate that directional detection is possible for a wide range of dark matter masses, and pave the way for light dark matter discovery with bulk superconducting targets.
\end{abstract}

\maketitle

The identity of dark matter (DM) remains one of the most pressing open questions in particle physics and cosmology. Contrary to decades of theoretical expectations, numerous experimental probes have found no conclusive evidence of DM at the weak scale, leading to renewed interest in models of DM at much lower scales. Many new ideas have recently been proposed to search for such light DM in the laboratory~\cite{Essig:2011nj,Graham:2012su,Essig:2015cda,Lee:2015qva,Hochberg:2015pha,Hochberg:2015fth,Alexander:2016aln,Derenzo:2016fse,Hochberg:2016ntt,Kavanagh:2016pyr,Emken:2017erx,Emken:2017qmp,Battaglieri:2017aum,Essig:2017kqs,Cavoto:2017otc,Hochberg:2017wce,Essig:2018tss,Emken:2018run,Ema:2018bih,Geilhufe:2018gry,Baxter:2019pnz,Essig:2019xkx,Emken:2019tni,Hochberg:2019cyy,Trickle:2019nya,Griffin:2019mvc,Coskuner:2019odd,Geilhufe:2019ndy,Catena:2019gfa,Blanco:2019lrf,Kurinsky:2019pgb,Kurinsky:2020dpb,Griffin:2020lgd,Radick:2020qip,Gelmini:2020xir,Trickle:2020oki,Du:2020ldo,Blanco:2021hlm}, and several of these novel direct detection exeperiments have already begun to probe significant parameter space~\cite{Essig:2012yx,Tiffenberg:2017aac,Romani:2017iwi,Crisler:2018gci,Agnese:2018col,Agnes:2018oej,Settimo:2018qcm,Akerib:2018hck,Abramoff:2019dfb,Aguilar-Arevalo:2019wdi,Aprile:2019xxb,Barak:2020fql,Arnaud:2020svb,Amaral:2020ryn}. Among the new ideas, superconducting targets stand out with the lowest possible thresholds, giving them sensitivity to the lowest DM masses through DM--electron interactions~\cite{Hochberg:2015pha,Hochberg:2015fth,Hochberg:2019cyy,Hochberg:2021pkt}. With superconducting energy gaps of $\mathcal O(\SI{}{\milli\electronvolt})$, such detectors may eventually probe DM with mass as low as the keV scale, where cosmological constraints become significant \cite{Tremaine:1979we,Boyarsky:2008ju,Boyarsky:2008xj}.

Despite the impressive potential reach of superconducting targets, current projections assume that the detectors are insensitive to the \textit{direction} of incoming DM. Directional detection has long been recognized as a powerful tool in DM experiments, including those in the keV--GeV regime \cite{Blanco:2021hlm,Coskuner:2019odd}: due to the halo wind, the local DM distribution is not isotropic in the laboratory frame, leading to a characteristic modulation of the signal that can be used to reject backgrounds and confirm a discovery. If superconducting detectors can be made sensitive to the direction of the incoming DM, then such targets will offer exceptional promise for future experiments. Such a detector would be capable of making a definitive discovery of DM as light as a keV, should a signal be observed.

In this work, we show that even isotropic superconductors are capable of directional detection via the angular distribution of the excitations produced by DM scattering. For such a measurement to be viable, two key features are required. Firstly, the direction of the initial excitations produced by the DM interaction should be correlated with that of the incoming DM particle. Secondly, the secondary excitations produced by the initial excitations as they down-convert in the material should exhibit directionality correlated with that of the initial excitations. As we will show, both features indeed occur in superconducting targets, paving the way for directional detection of keV-scale DM.

This Letter is organized as follows. We begin by considering the initial scattering of DM with electronic states of a superconductor into excited quasiparticles. Next we consider the down-conversion of these initial excitations in the material into secondary quasiparticles and phonons, treating the general case with a new numerical code. We then present our results for directionality, and end with a discussion of experimental prospects.

Throughout this work, we use the following notation: for a 3-vector $\bb q$, we write $q = \left|\bb q\right|$. We use angles with two subscripts to denote the relative angle between the two axes specified by those subscripts. All other angles are defined relative to the DM wind axis. The Fermi energy and momentum are denoted by $E_\fermi$ and $p_\fermi$, respectively. We use natural units where $\hbar = c = 1$.

\section{Dark Matter Scattering}
We begin by demonstrating the directionality of the initial excitations produced in a DM scattering event in a superconducting target. This requires  reformulating the description of the scattering process in terms of the appropriate degrees of freedom in the `BCS' vacuum of the superconductor \cite{Bardeen:1957mv}. The DM scattering rate in superconductors was originally computed by \refcite{Hochberg:2015fth} considering only large energy deposits compared to the superconducting gap, and so the DM--detector interaction was described in terms of the DM $\dm$ scattering with individual electrons, $\left|\dm\right\rangle\left|e^-\right\rangle\longrightarrow\left|\dm'\right\rangle\left|e^-\right\rangle$. While this description is appropriate for computing the rate, it is not suitable for studying the kinematics at small deposits. Here, the appropriate degrees of freedom are Bogoliubov quasiparticles (QPs)~\cite{Bogolyubov:1958se}, which are electron--hole superpositions.

In this description, the DM excites the BCS vacuum by pair-producing QPs, as $\left|\dm\right\rangle\left|0_\BCS\right\rangle \longrightarrow \left|\dm'\right\rangle\left|\qp_1, \qp_2\right\rangle$. The total momentum of these \textit{two} QPs is the momentum transfer $\bb q$ imparted by the DM scatter. The wave functions of the electrons in the BCS vacuum automatically account for Pauli blocking through a \textit{coherence factor,} which has significant support only when one of the QPs is below the Fermi momentum and the other is above. In the Supplemental Material (SM), we show that for energy deposits much larger than the superconducting gap energy, the scattering rate becomes identical to that for scattering with individual electrons: the matrix element for QP pair production becomes that for electron scattering, the coherence factor becomes a Pauli blocking factor, and the two QPs become an electron--hole pair.
We use the labels $\qpa$ and $\qpb$ to refer to the two initial QPs produced in the scattering process. All quantities are invariant under interchange of QPs $\qpa$ and $\qpb$, so we use the label $\qpi$ for statements that apply to either label.

The QPs have a dispersion relation different from that of free electrons, of the form 
\begin{equation}
    \label{eq:dispersion}
    E_{\qp}({\bb p}) = \sqrt{
        \Erelative_{\bb p}^2
        + \Delta^2
    }
    \,,
\end{equation}
where $\Erelative_{\bb p} = \hbarsq\bb p^2/(2\mstar)-E_{\fermi}$ is the Bloch energy relative to the Fermi surface and $\Delta$ is half of the superconducting gap energy. Counterintuitively, the energy of a QP is minimized ($E_\qp = \Delta$) when its momentum lies \textit{on} the Fermi surface. The free-electron dispersion relation is recovered in the limit $p_\qp\gg p_\fermi$, whereas the limit $p_\qp\ll p_\fermi$ gives the energy of a hole far below the Fermi surface. This non-trivial dispersion relation modifies the kinematics of DM scattering near the gap, and thus significantly influences directional correlations and down-conversion. For energy deposits $\myhbar\omega \lesssim \SI{}{\kilo\electronvolt}$, the momenta $\bb p_1$ and $\bb p_2$ will be well inside the first Brillouin zone (BZ). For deposits $\omega \gtrsim \SI{}{\kilo\electronvolt}$, we expect \cref{eq:dispersion} to receive band structure corrections near the edge of a BZ of order tens of eV or less, small compared to this keV scale, so \cref{eq:dispersion} is a valid approximation for all the energy scales considered in this work.

The overall DM scattering rate is given by \cite{Hochberg:2021pkt}
\begin{equation}
    \label{eq:overall-rate}
    \Gamma(v_\dm) = {\myhbar}\overall\int
    \frac{\du^3\bb q}{(2\pi)^3}
    \left|V(\bb q)\right|^2\!\frac{2\bb q^2}{e^2}
    \Im\left[-\frac{1}{\epsilon_\BCS(\bb q,\omega_{\bb q})}\right]
    ,
\end{equation}
where $v_\dm = |\bb v_\dm|$ is the magnitude of the DM velocity; $\bb q$ is the momentum transfer; $\myhbar\omega_{\bb q} = \bb q\cdot\bb v_\dm - \bb q^2/2m_\dm$ is the deposited energy; and $\epsilon_\BCS$ is the dielectric function of a superconductor in the BCS vacuum. In this work, we make the approximation $\Im\left(-1/\epsilon_\BCS\right) \equiv \Im\left(\epsilon_\BCS\right) / \left|\epsilon_\BCS\right|^2 \simeq \Im\left(\epsilon_\BCS\right)/|\epsilon_\lindhard|^2$. Here   $\epsilon_\lindhard$ is the Lindhard form of the dielectric function for a normal metal~\cite{dressel2002electrodynamics}, which accounts for the effects of in-medium screening and collective modes in the normal metal phase~\cite{Hochberg:2021pkt}. We compute $\Im\left(\epsilon_\BCS\right)$ in terms of the QP dispersion relation \cref{eq:dispersion} and the BCS coherence factor, which accounts for near-gap effects \cite{tinkham:2004} and for Pauli blocking. Our approach interpolates between the approximate superconducting dielectric response near the gap and the normal-metal response far from the gap. A more complete treatment explicitly computing the dielectric function in the BCS vacuum lies beyond the scope of this work and will pursued elsewhere~\cite{futureBCS}. We take $\left|V(\bb q)\right|^2 = (g_eg_\dm)^2\bigl(\bb q^2  + m_\med^2 \csq\hbarsq\bigr)\null^{-2}$, where $g_e$ and $g_\dm$ are the couplings of the mediator to the electron and the DM, respectively. This is appropriate for any spin-independent interaction. Further details on DM interactions and the computation of $\Im\left(\epsilon_\BCS\right)$ are given in the SM.

The rate in \cref{eq:overall-rate} can be differentiated to obtain the parameter distributions of the initial QPs prior to down-conversion. We begin with the joint distribution of $\costh_{\dm\qpa}$, $\costh_{\dm\qpb}$, $E_\qpa$, and $E_\qpb$, where $\theta_{\dm\qpi}$ is the angle between $\bb p_\dm$ and $\bb p_\qpi$. While the distribution of the $\costh_{\dm\qpi}$ fully describes the directionality of the initial excitations, the energies of these excitations are also needed to determine the properties of secondary excitations. The four-parameter joint distribution $f_\qp[\bb p_\dm](\costh_{\dm\qpa}, \costh_{\dm\qpb}, E_\qpa, E_\qpb)$ at fixed $\bb p_\dm$ can be combined with a DM distribution $f_\dm(\bb p_\dm)$ to obtain the full distribution $f_\qp(\costh_\qpa, \costh_\qpb, E_\qpa, E_\qpb)$, where $\theta_\qpi$ is the angle between the QP momentum and the DM \textit{wind} axis. In our numerical results, we draw samples from this distribution to generate initial excitations.

\begin{figure*}
    \centering
    \includegraphics[width=2\columnwidth]{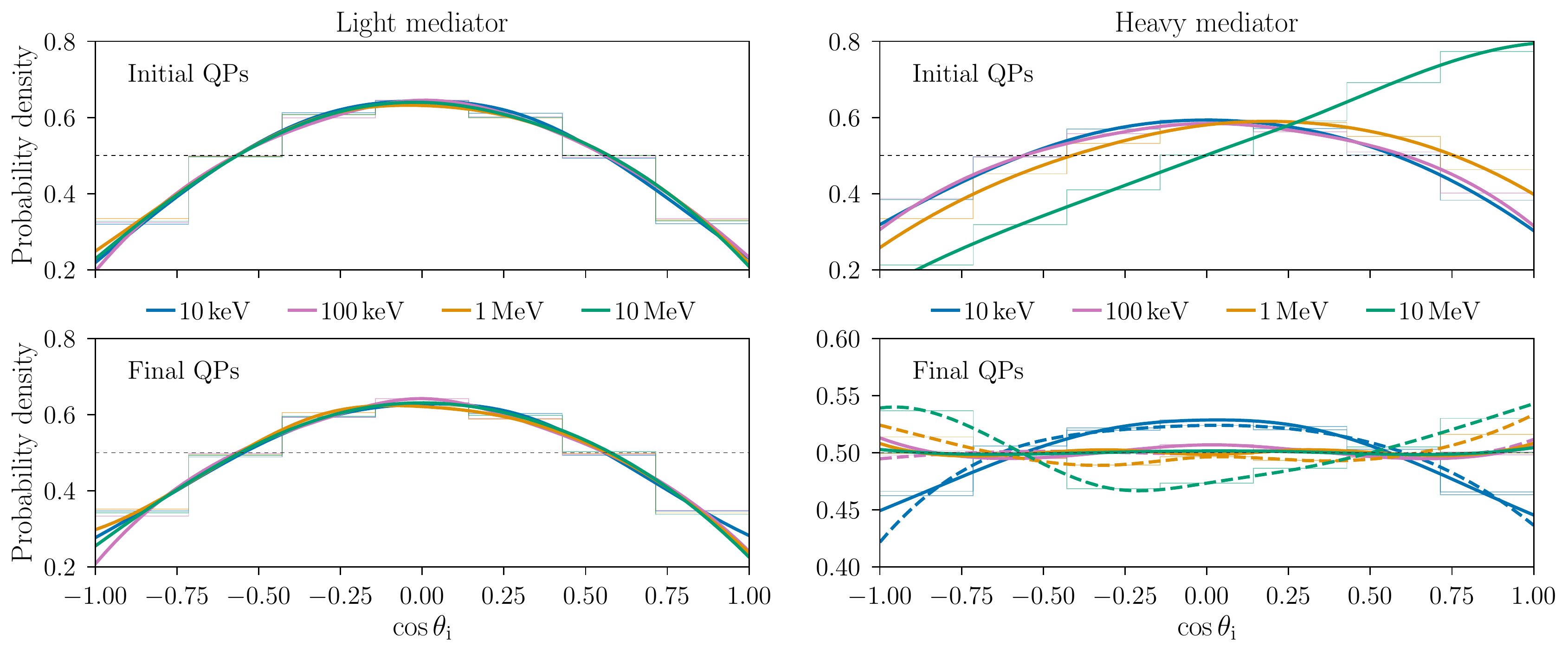}
    \caption{Angular distributions of QPs produced by DM scattering in Al. The angles shown are defined with respect to the axis of the DM wind. The distribution of DM orientations in the Standard Halo Model is included. The left and right column show the distributions in the light- and heavy-mediator limits, respectively. In each panel, the colors correspond to different DM masses, and a dashed horizontal line at $\costh_\qpi = \frac12$ indicates the isotropic distribution. Thick lines interpolate between histogram values (thin lines) for ease of visualization. The top and bottom rows show distributions of QPs before and after down-conversion, respectively. In the bottom-right panel, several solid curves overlap near the isotropic distribution. The dashed curves show angular distributions obtained by restricting to events with total deposit $\omega < 20\Delta$, for which the effects of down-conversion are less significant.}
    \label{fig:angular-distribution-summary}
\end{figure*}

In certain regimes, the angular distribution of excitations can be understood analytically by virtue of kinematical constraints. Conservation of energy yields a closed-form expression for $\costh_{\bb q\qpi}$, the cosine of the angle between QP $\qpi$ and the momentum transfer $\bb q$, in terms of $q$ and $p_\qpi$. Now consider small DM masses and light mediators, where small deposits are favored. In the limit of small deposits, to leading order in $\omega-2\Delta$, we have $p_\qpi \simeq p_\fermi$. Then conservation of momentum requires
\begin{equation}
    \costh_{\bb q\qpi} \simeq
        \frac{m_\dm v_\dm
            - \sqrt{2m_\dm\left(\frac12m_\dm v_\dm^2 - 2\Delta\right)}}
        {2\mstar v_\fermi}
    \;,
\end{equation}
where $v_\chi=|\bb v_\chi|$, which implies $0 \lesssim \costh_{\bb q\qpi} \lesssim 2\Delta/(v_\dm p_\fermi)$. For aluminum (Al), with $\Delta = \SI{0.3}{\milli\electronvolt}$ and $E_\fermi = \SI{11.7}{\electronvolt}$, this leads to the condition $0\lesssim\costh_{\bb q\qpi}\lesssim\num{e-4}$, so excitations produced near the gap are nearly orthogonal to the momentum transfer $\bb q$. In turn, the direction of $\bb q$ is necessarily correlated with that of $\bb v_\dm$, so the distribution of $\costh_{\dm \qpi}$ is peaked at zero.

On the other hand, consider DM interacting via a heavy mediator, for which large deposits are favored. In particular, for $p_\dm \sim p_\fermi$, momentum transfers of order $p_\fermi$ are possible, corresponding to $\costh_{\bb q\qpi} \sim 1$. For example, for $p_\dm = 2p_\fermi$, if the DM is fully stopped and its energy shared equally between the two QPs, then $\costh_{\bb q\qpi}$ is given uniquely by
$\costh_{\bb q\qpi} = \sqrt{v_\fermi/(v_\fermi + v_\dm)}$.
For typical materials, $v_\dm \ll v_\fermi$, so indeed $\costh_{\bb q\qpi} \approx 1$. In Al, this solution corresponds to $\costh_{\bb q\qpi} \approx 0.93$. Further, fully stopping the DM implies that $\costh_{\dm\bb q} = 1$, so $\costh_{\dm\qpi} = \costh_{\bb q\qpi}$. Thus, when $p_\dm \sim p_\fermi$ and large $q$ is favored, the angular distribution can peak in the direction of the DM wind.

The marginal distribution of $\costh_\qpi$ is shown by the solid curves in the top panels of \cref{fig:angular-distribution-summary} for several DM masses in Al. We assume the Standard Halo Model (SHM), \textit{i.e.,} $f_\dm(v) \propto \Theta(v_{\mathrm{esc}} - v)e^{-v^2/v_0^2}$ in the galactic frame, taking $v_0 = \SI{220}{\kilo\meter/\second}$, $v_{\mathrm{esc}} = \SI{550}{\kilo\meter/\second}$, and Earth velocity $v_{\mathrm{E}} = \SI{230}{\kilo\meter/\second}$ relative to the galactic frame. For light DM or a light mediator, small energy transfers are favored, leading to a peak in the angular distribution orthogonal to the DM wind axis. For heavier DM and mediators, larger energy transfers lead to a forward-peaked distribution.

\section{Quasiparticle relaxation}
The scattering process described above produces a single pair of QPs with a known total energy and net momentum, and we showed how the directions of these quasiparticles are related to the direction of the incoming DM. The second requirement for directional detection is that this directionality must be preserved after the initial excitations relax. Thus, we now study the down-conversion of the initial QP excitations through QP--phonon scattering.

Following \refscite{gregory,cs,bardeen}, we model down-conversion as a repeating sequence of two distinct processes: first, energetic QPs relax by emission of phonons, and second, energetic phonons decay into QP pairs. Quasiparticle pair production eventually stops once all remaining phonons have energy below $2\Delta$. We treat such phonons as ballistic, including them as part of the final state that is eventually read out by the detector. As shown in the SM, phonon emission is kinematically forbidden for very low-energy QPs, so the QPs eventually become ballistic as well. The down-conversion process is finished when all particles are ballistic. For other approaches to the relaxation of highly energetic QPs, see \refscite{kozorezov,OK}.

We study the impact of down-conversion on directionality by explicit simulation, implemented in a public code released together with this Letter.\footnote{\url{http://github.com/benvlehmann/scdc}} We begin with an ensemble of initial excitations sampled from the distribution $f_\qp$, and then iterate the relaxation processes described above until all QPs and phonons are ballistic. Computing the momentum of the outgoing excitations after each relaxation process requires knowledge of the differential rate of these processes with respect to the kinematical parameters of the final state. We take the differential rates for phonon emission and QP pair production from Eqs.~(16) and (27) of \refcite{Kaplan:1976zz}. In each case, imposing conservation of energy and momentum using the dispersion relation of \cref{eq:dispersion} gives the distribution of final-state angles. The differential rate of QP pair production by a phonon of energy $\omegaph_\ph$ is given by
\begin{equation}
    \label{eq:qp-pair-distribution}
    \frac{\du\Gamma}{\du E_\qpi}
    \propto
    \frac{E_\qpi (\omegaph_\ph - E_\qpi) + \Delta^2}{
        \sqrt{\bigl[E_\qp^2 - \Delta^2\bigr]\bigl[
                (\omegaph_\ph - E_\qpi)^2 - \Delta^2
            \bigr]
        }
    }
    \,,
\end{equation}
where $E_\qpi$ is the energy of one QP and $\omegaph_\ph - E_\qpi$ is the energy of the other. This distribution is sharply peaked at $E_\qpi\sim\Delta$ and $E_\qpi\sim \myhbar\omega_\ph - \Delta$, corresponding to the case in which one of the two QPs receives most of the phonon's energy. Here, the dispersion relation of \cref{eq:dispersion} implies that the QPs are produced nearly orthogonal to the axis of the initial phonon. Thus, for small deposits, the angular distributions of QPs and phonons in the final state will peak in orthogonal directions.

We sample initial excitations and simulate the down-conversion process for several DM masses in Al. The angular distributions of the resulting QPs are shown as the solid curves in the bottom panels of \cref{fig:angular-distribution-summary}. The angular distribution of phonons exhibits weaker directionality and is discussed in the SM. We learn that for light mediators, the initial QPs have excellent directionality, which is well-preserved after down-conversion, as the initial and down-converted distributions are comparable. For a heavy mediator, despite the directionality of the initial QPs, the down-converted distribution is much closer to isotropic, particularly for heavier DM. This is simply because heavy mediators and heavy DM favor larger deposits, which lead to a larger number of relaxation events. The directional correlation of the final states with the DM axis is reduced with each relaxation event, so directionality is best preserved when $\myhbar\omega/\Delta - 2$ is small. For this reason, the dashed curves in the bottom-right panel of \cref{fig:angular-distribution-summary} show the angular distribution resulting from including only events with total deposits $\myhbar\omega < 20\Delta$. These dashed curves retain directionality associated with the low-energy part of the initial angular distribution.

\section{Results}

\begin{figure*}\centering
    \includegraphics[width=\columnwidth]{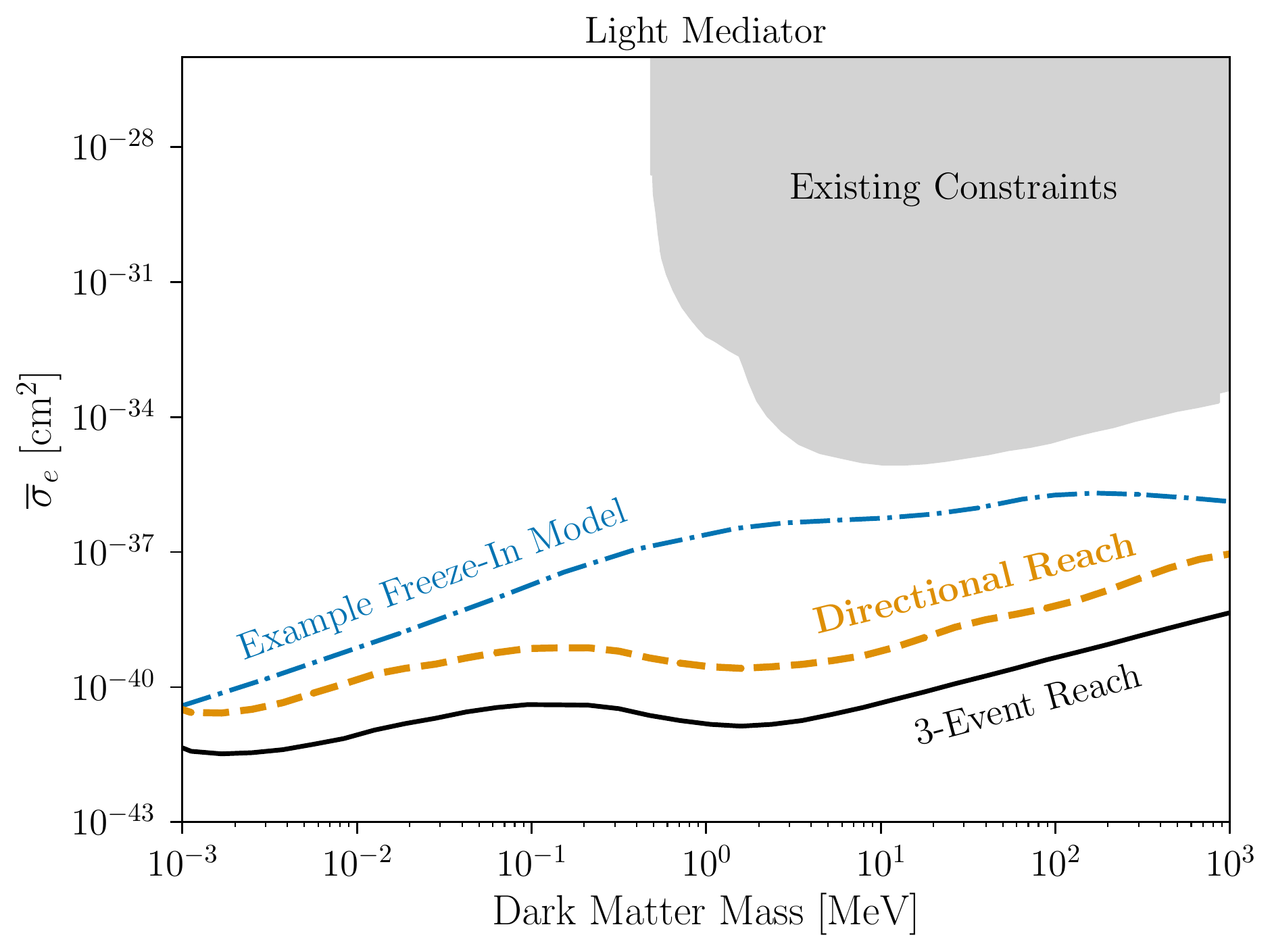}
    \hfill
    \includegraphics[width=\columnwidth]{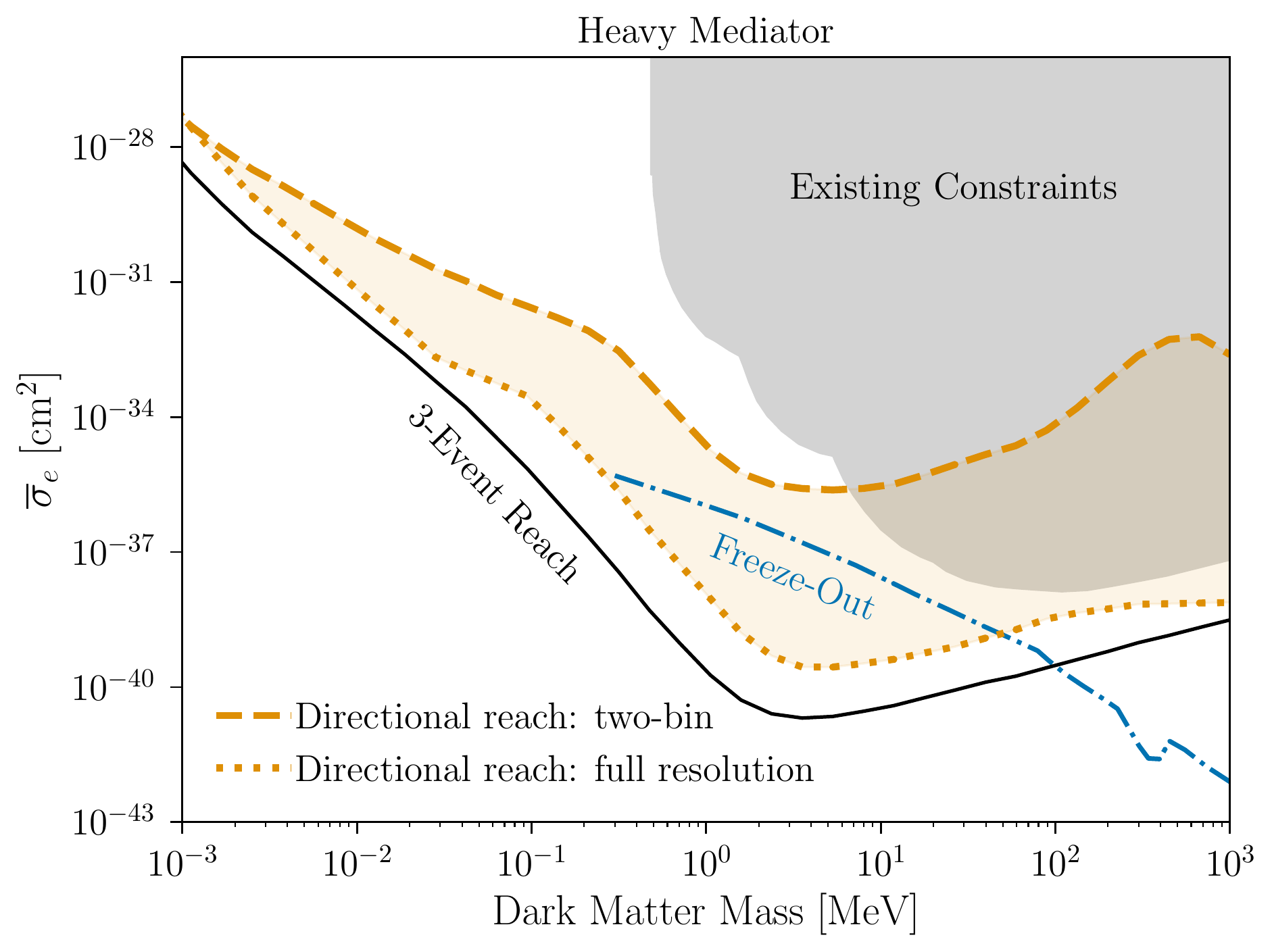}
    \caption{Directional detection discovery reach for DM scattering in an Al superconductor via a light (\textit{left panel}) or heavy (\textit{right panel}) mediator.  Solid lines: 3-event reach for a kg-yr exposure, not including directionality. Dashed lines: estimated discovery reach for directionality at 95\% C.L. using only two angular bins. Dotted line: estimated discovery reach for directionality in an experiment with high-precision measurement of $\costh$.
    Blue dot-dashed lines show cross sections for example DM models \cite{Battaglieri:2017aum}.
    Shaded gray regions indicate existing constraints from SENSEI~\cite{Barak:2020fql}, SuperCDMS HVeV~\cite{Amaral:2020ryn}, DAMIC~\cite{Aguilar-Arevalo:2019wdi}, Xenon10~\cite{Essig:2017kqs}, DarkSide-50~\cite{Agnes:2018oej}, and Xenon1T~\cite{Aprile:2019xxb}.}
    \label{fig:reach}
\end{figure*}

\Cref{fig:reach} shows the estimated sensitivity for the detection of the directional DM wind, with impressive reach. Critically, since the target is \textit{isotropic}, there is no modulation in the overall rate, unlike many directional detection schemes. Instead, the reach must be defined in terms of the anisotropy in the distribution of the final-state excitations. The dashed lines in \cref{fig:reach} show the projected reach for a detector which counts final-state quasiparticles in each of two angular bins: the `on-axis' bin, with $\left|\costh\right| > \frac12$, and the `off-axis' bin, with $\left|\costh\right| < \frac12$. Such a counting experiment requires a certain number of events in order to establish that the signal is not isotropic at a certain confidence level.\footnote{As a benchmark, if the signal falls entirely in one of the two bins, six events are required to rule out an isotropic distribution at 95\% C.L.} In turn, this number of events translates to a minimal cross section for a fixed experimental exposure. Note that for a heavy mediator, we impose a cut to include only small deposits, which reduces the overall rate, but lessens the impact of down-conversion.

The two-bin configuration is the minimal experimental configuration for directional detection, and represents the most conservative projection. For heavy mediators, a more ambitious projection is obtained by assuming precise resolution of $\costh$, as well as sufficient time resolution so that final-state quasiparticles can be grouped by event. In this case, the detailed shape of the angular distribution can be compared with the distribution under the null hypothesis of an isotropic background. This again translates to a minimal number of events to detect directionality, and a corresponding minimal cross section, shown by the dotted line in \cref{fig:reach}. Further details are given in the SM. For light mediators, since directionality is manifest, this procedure gives almost exactly the same result as the two-bin configuration, so the corresponding line is not shown.

While these are both schematic estimates, the dashed and dotted curves in \cref{fig:reach} are representative of the directional sensitivity of an experiment in the simplest and most sophisticated configurations. \Cref{fig:reach} shows that a kg-yr exposure of such an experiment would be capable of detecting directionality for DM masses $\SI{}{\kilo\electronvolt}\lesssim m_\dm \csq \lesssim \SI{}{\giga\electronvolt}$, throughout the discovery range of superconducting detectors. Both configurations discussed here would be directionally sensitive at cross sections covering important cosmological targets that are not currently probed by any direct detection experiments.

\section{Discussion}
In this work, we have shown that the parameter space to be accessed by superconducting targets~\cite{Hochberg:2015pha,Hochberg:2015fth,Hochberg:2021pkt} can be probed \textit{directionally}, even in the case of an isotropic medium. This has important implications for the real-world design of directional superconducting detectors. Directionality will require detectors to push resolutions lower, with thresholds close to the superconducting gap allowing the best coverage of DM parameter space. Detectors that can trap primary QPs are preferred as they will be able to take advantage of the directional correlations in the DM signal. This is in contrast to the weak directionality in the phonon system, in terms of total energy.

Importantly, in order to scale  detectors to kg-yr exposures while retaining directional sensitivity,  a massive multiplexing scheme will likely be required. As summarized in \refcite{Hochberg:2015fth}, typical detector volumes will be of order \SI{1}{\centi\meter^3} or smaller in order to attain high collection efficiencies. Mean QP lifetimes are discussed in more detail in \refcite{Kaplan:1976zz}, and intrinsic lifetimes (even in the `dirty' limit) do diverge for very low temperatures and QP occupancy. 

For realistic applications of bulk superconducting targets, characterization of mean QP diffusion length in real samples---which will probe the effects of impurities and lattice defects on the scattering lifetime of otherwise ballistic QPs---will determine which materials are best suited for directional DM detection. Such work has been done for large niobium (Nb) crystals \cite{gaitskell1993a}, but has largely been put aside over the last few decades. These programs will need to be restarted in order to characterize samples with the appropriate properties to detect small energy deposits in the regime relevant to directional DM detection. For materials with a known diffusion length, this directionality can be converted to a rate modulation by making a detector in which one path length to the sensor is much shorter than this diffusion length, and the orthogonal path length is much longer than the diffusion length. Chemical Vapor Deposition--grown superconducting crystals of Nb or Al instrumented on their large surface, with cross-sections of a few \SI{}{\milli\meter^2} and thickness of around 100 microns, would achieve this behavior for typical diffusion lengths of a few hundred microns.

On the theoretical side, our results demonstrate directional detection of DM in a target that is otherwise isotropic, in contrast to most directional studies which utilize anisotropy in the material response or in the geometry of the experiment. Here, directionality is inferred from the geometric properties of the excitations themselves, rather than from a rate variation. Our results strongly suggest that for a gapless material with typical acoustic phonon modes, no directional correlation is preserved between the initial DM scatter and the outgoing excitations, as phonons can always be emitted in the limit $\Delta\rightarrow 0$. In an anisotropic material, some correlations may still persist further above the gap due to an increased number of forbidden transitions. As an example, indirect-gap materials, if such superconductors exist, would be much more likely to preserve directionality even in the limit of a small gap if large energies are required for inter-valley transitions. We leave the exploration of such scenarios for future work~\cite{future}, as they require detailed study of the anistropy in the DM response as well.

\medskip

\textbf{Acknowledgments.}
We thank Roni Ilan and Adolfo Grushin for many useful conversations about superconductors, and Kathryn Zurek for discussions at early stages of this work. YH is grateful to Nadav Katz for a conversation in the Ross dungeon that led to this work. We thank Yonatan Kahn for comments on a draft version of this manuscript.
The work of YH is supported by the Israel Science Foundation (grant No. 1112/17), by the Binational Science Foundation (grant No. 2016155), by the I-CORE Program of the Planning Budgeting Committee (grant No. 1937/12),  and by the Azrieli Foundation. EDK was supported by the Zuckerman Foundation and by the Israel Science Foundation (grant No. 1111/17). Parts of this document were prepared by NK using the resources of the Fermi National Accelerator Laboratory (Fermilab), a US Department of Energy, Office of Science, HEP User Facility. Fermilab is managed by Fermi Research Alliance, LLC (FRA), acting under Contract No. DE-AC02-07CH11359. The work of BVL is supported in part by DOE grant DE-SC0010107. 

\bibliographystyle{JHEP}
\bibliography{references}

\onecolumngrid
\clearpage

\setcounter{page}{1}
\setcounter{equation}{0}
\setcounter{figure}{0}
\setcounter{table}{0}
\setcounter{section}{0}
\setcounter{subsection}{0}
\renewcommand{\theequation}{S.\arabic{equation}}
\renewcommand{\thefigure}{S\arabic{figure}}
\renewcommand{\thetable}{S\arabic{table}}
\renewcommand{\thesection}{\Roman{section}}
\renewcommand{\thesubsection}{\Alph{subsection}}
\newcommand{\ssection}[1]{
    \addtocounter{section}{1}
    \section{\thesection.~~~#1}
    \addtocounter{section}{-1}
    \refstepcounter{section}
}
\newcommand{\ssubsection}[1]{
    \addtocounter{subsection}{1}
    \subsection{\thesubsection.~~~#1}
    \addtocounter{subsection}{-1}
    \refstepcounter{subsection}
}
\newcommand{\fakeaffil}[2]{\textsuperscript{#1}\textit{#2}\\}

\thispagestyle{empty}
\begin{center}
    \begin{spacing}{1.2}
        \textbf{\large 
            Supplemental Material:\\Directional Detection of Light Dark Matter in Superconductors
        }
    \end{spacing}
    \par\smallskip
    Yonit Hochberg,\textsuperscript{1}
    Eric David Kramer,\textsuperscript{1}
    Noah Kurinsky,\textsuperscript{2,~3,~4}
    and Benjamin V. Lehmann,\textsuperscript{5,~6}
    \par
    {\small
        \fakeaffil{1}{Racah Institute of Physics, Hebrew University of Jerusalem, Jerusalem 91904, Israel}
        \fakeaffil{2}{Fermi National Accelerator Laboratory, Batavia, IL 60510, USA}
        \fakeaffil{3}{Kavli Institute for Cosmological Physics, University of Chicago, Chicago, Illinois 60637, USA}
        \fakeaffil{4}{SLAC National Accelerator Laboratory, Menlo Park, CA 94025}
        \fakeaffil{5}{Department of Physics, University of California Santa Cruz, Santa Cruz, CA 95064, USA}
        \fakeaffil{6}{Santa Cruz Institute for Particle Physics, Santa Cruz, CA 95064, USA}
        (Dated: \today)
    }
\end{center}
\par\smallskip

In this Supplemental Material, we first present the details of quasiparticle (QP) production by dark matter (DM) scattering. We then provide a pedagogical discussion of QP down-conversion in superconductors, supply derivations of the directionality results in the main text, and compare the results to the normal metal case. Finally, we briefly demonstrate the dependence of post--down-conversion directionality on the energy deposited, and show the resulting asymmetry in the recoil spectra.

\ssection{Dark Matter Interactions}
In this section, we detail our treatment of DM--electron interactions in the language of QP pair production. In particular, we compute $\Im\epsilon_\BCS$ using the BCS coherence factor $\FBCS(p_\qpa, p_\qpb)$, and we show that the free electron scattering picture is recovered in the limit of large deposits. Throughout this section, $q$ denotes the 4-momentum transfer $(\omega, \bb q)$.

Consider a DM--electron interaction mediated by a scalar particle $\varphi$, with interaction Lagrangian
\begin{equation}
    \mathcal{L}_{\rm int} =
        g_\dm \phi\,\overline\dm\dm + g_e \phi\, \overline \psi \psi
    \;,
\end{equation}
where $\dm$ is a spin-$1/2$ DM fermion and $\psi$ is the electron. Using the projection operator $P=(1+ \gamma^0)/2$, we can project out the so-called `large part' \cite{Bjorken:1965zz} of the electron field $\psi_s$, where $s=\uparrow,\downarrow$ refers to the two different spin states. This gives, at lowest order, the interaction Hamiltonian for the electron--mediator interaction:
\begin{equation}
\label{eq:Hint}
    H_{\mathrm{int}}
    \simeq
    -g_e\int\du^3{\bb x}\, \phi(\bb x)
    \Bigl(
        \psi^\dagger_\uparrow(\bb x)\psi{}_\uparrow(\bb x)
        + \psi^\dagger_\downarrow(\bb x)\psi^{}_\downarrow(\bb x)
    \Bigr)
    \;.
\end{equation}
The DM--mediator interaction is governed by a Hamiltonian of the same form with the replacement $\psi \to \dm$. Higher-order terms are discussed by \refcite{Mitridate:2021ctr}. (Similarly, as shown in that reference, for a light vector mediator $A^\mu$, we will have in the low-energy limit $\mathcal{L}_{\rm int} = g_\dm A_0 \dm^\dagger \dm + g_e A_0 \psi^\dagger \psi$. Higher-order terms and magnetic interactions will be suppressed by factors of $v_\dm/c$ in the low-energy limit. In the case of a heavy vector mediator, the $A_0$ interaction will again dominate, because the currents in the interaction $\bb A\cdot \bb j$ will be suppressed by factors of $v_\dm/c$. Thus, light and heavy vector mediators should also be described by interaction \cref{eq:Hint} in this limit.\ignorespaces
\footnote{\ignorespaces
    \begin{minipage}[t]{0.99\textwidth}
        One might object that a propagating $A_0$ should be suppressed in the non-relativistic limit, by virtue of the constraint $\del_\mu A^\mu=0$. Indeed, one can compute the time-ordered propagator $\int\!\du { t}\, \du^3\bb x\,\exp({iq^0 t - i\bb q\cdot \bb x})\langle 0|T\{A^\mu(\bb x,{ t}),A^\nu(\bb 0,0)\}|0\rangle$ in the interaction picture and verify that it is not Lorentz-covariant, and that its 00 component is highly suppressed when $|\bb q|\ll m_A$. However, the absence of a kinetic term for $A_0$ in the Lagrangian introduces an additional terms in the Hamiltonian to precisely cancel this suppression. The effective propagator becomes the Lorentz-covariant propagator $-i/(q^2-m^2_A+i\varepsilon)(\eta^{\mu\nu}-q^\mu q^\nu/m^2_A)$, which is unsuppressed for $q^0\ll |\bb q|$, or when coupling to a conserved current. This is equivalent to the statement that an off-shell vector can be polarized in any direction. See Sec.~6.2 of \refcite{Weinberg:1995mt} for further insight.
    \end{minipage}\ignorespaces
})
Defining the density operator $\rho_e(\bb x) \equiv \sum_s \psi^\dagger_s(\bb x)\psi_s(\bb x)$, its Fourier transform is
\begin{equation}
    \rho_e(\bb q) =
        \sum_s \int
        \frac{\du^3{\bb p}}{(2\pi)^3}c^\dagger_{\bb p-\bb q,s}c_{\bb p,s}
    ,
\end{equation}
where $c_{\bb p,s}$ annihilates an electron with momentum $\bb p$ and spin $s$. A similar expression is obtained for $\rho_\dm$. We can then write the interaction Hamiltonian as
\begin{equation}
    H_{\mathrm{int}} = 
    -\int\!\du^3{\bb x}\,
    \varphi(\bb x)\Bigl(g_e\rho_e(\bb x) +g_\dm\rho_\dm(\bb x)\Bigr)
    \;.
\end{equation}
At second order in perturbation theory, the $S$-matrix will therefore contain a term
\begin{align}
    \label{eq:smatrix}
    \hat{\mathbb{S}}^{(2)}
    &\supset
    -g_e g_\dm \!\int\!\du^4x\dd^4x'\,
        \overline\dm(x) \dm(x) \Delta(x-x')
        \overline\psi(x') \psi(x')
    \nonumber\\
    &=
    -{g_eg_\dm}\int\!\frac{\du^4q}{(2\pi)^4}\;
        i\frac{\rho_\dm^\dagger(q)\,
        \rho_e(q)}{q^2-m_\phi^2+i\varepsilon}
    \;,
\end{align}
where
\begin{equation}
    \label{eq:rhoq4}
    \rho_e(q)\equiv\rho_e(\bb q,\omega)
    = \int\!\du t\,e^{i\omega t}\rho_e(\bb q,t)
    = \int\!\du t\,e^{i\omega t}e^{iH_0t}\rho_e(\bb q)\,e^{-iH_0 t}\;,
\end{equation}
with $H_0$ the free Hamiltonian. In the presence of the lattice potential, an effective electron-electron potential is induced through a phonon loop \cite{Polchinski:1992ed}. The energy eigenstates in the presence of this effective potential are now given by the Bogoliubov QPs, with creation/annihilation operators $\gamma^\dagger$, $\gamma$ respectively \cite{Bogolyubov:1958se}. To implement the unitary transformation to the QP basis, we simply replace
\begin{equation}
    c_{\bb p\uparrow} = u_{\bb p} \gamma_{\bb p\uparrow}
        + v_{\bb p} \gamma^\dagger_{-\bb p,\downarrow}
    ,
    \hspace{1cm}
    c^\dagger_{-\bb p,\downarrow} = - v_{\bb p} \gamma_{\bb p\uparrow}
        + u_{\bb p} \gamma^\dagger_{-\bb p\downarrow}
    ,
\end{equation}
where the coefficients $u_{\bb p}$ and $v_{\bb p}$ satisfy
\begin{equation}
    |u_{\bb p}|^2 =
        \frac12\left(1+\frac{\Erelative_{\bb p}}{E_{\qp}(\bb p)}\right)
    ,
    \hspace{1cm}
    |v_{\bb p}|^2 =
        \frac12\left(1-\frac{\Erelative_{\bb p}}{E_{\qp}(\bb p)}\right)
    \;,
\end{equation}
with $\Erelative$ and $E_{\qp}$ defined as in \cref{eq:dispersion}. We can then isolate the term in $\rho_e$ that creates two quasiparticles (breaks a Cooper-pair):
\begin{equation}
    \rho_e(\bb q) \supset
    \int\!\frac{\du^3{\bb p}}{(2\pi)^3}
    (u^*_{\bb p+\bb q} v_{\bb p}+u_{\bb p}v^*_{\bb p +\bb q})
    \gamma^\dagger_{-\bb p -\bb q\uparrow}\gamma^\dagger_{\bb p \downarrow}
\end{equation}
giving, according to \cref{eq:rhoq4},
\begin{equation}
    \rho_e(\bb q,\omega) =
        \int\frac{\du^3\bb p}{(2\pi)^3}
        (u^*_{\bb p+\bb q} v_{\bb p}+u_{\bb p}v^*_{\bb p +\bb q})
        \gamma^\dagger_{-\bb p -\bb q\uparrow}
        \gamma^\dagger_{\bb p \downarrow}
        (2\pi)\delta\bigl(\omega-E_{\qp}(\bb p)-E_{\qp}(\bb p +\bb q)\bigr)
        \;.
\end{equation}

\begin{figure}\centering
    \includegraphics[width=0.55\columnwidth]{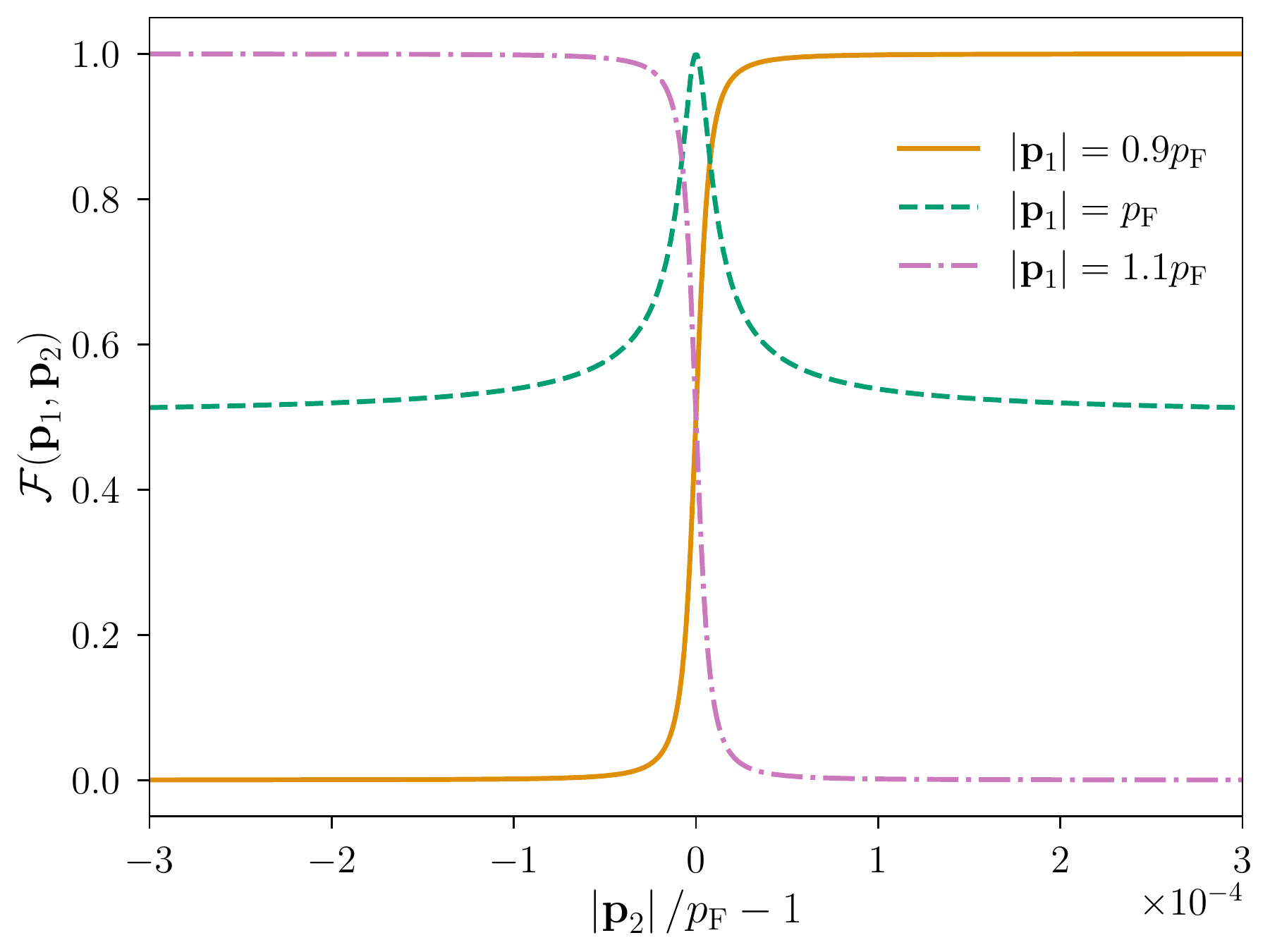}
    \caption{The BCS coherence factor, $\FBCS(\bb p_1, \bb p_2)$, for several fixed values of $|\bb p_1|$. Note that $\FBCS(\bb p_1, \bb p_2) = \FBCS(\bb p_2, \bb p_1)$. When both momenta are far from the Fermi surface, the coherence factor reduces to the Pauli blocking factor. Pauli blocking only permits the creation of an electron--hole pair if the electron is above the Fermi surface and the hole is below. Accordingly, if both of $\bb p_1$ and $\bb p_2$ are on the same side of the Fermi surface, the coherence factor vanishes rapidly. Otherwise, it quickly approaches 1. That $\FBCS(p_\fermi, p_\fermi) = 1$ is a consequence of the sign in the coherence factor for the interactions considered here. For interactions with the opposite sign, $\FBCS(p_\fermi, p_\fermi) = 0$.}
    \label{fig:coherence-factor}
\end{figure}

Plugging this into the S-matrix of \cref{eq:smatrix}, and using the fact that the scattering rate is given by $\Gamma = \frac{\du}{\du t} \sum_f|\langle f |\hat{\mathbb{S}}|i\rangle|^2$, we find that the lowest order Cooper-pair breaking rate is then simply given by
\begin{equation}
    \label{eq:rate}
    \Gamma(v_\chi) =  g^2_eg^2_\dm\int\!\frac{\du^3 \bb p_\qpa}{(2\pi)^3}
    \frac{\du^3\bb p_\qpb}{(2\pi)^3}\,
    2\pi\delta\Bigl(
        \omega_{\bb p_1 +\bb p_2}- E_\qp(\bb p_1) - E_\qp(\bb p_2)
    \Bigr)
    \frac{
        \bigl|u^*_{\bb p_\qpa} v_{\bb p_\qpb}
        +u_{\bb p_\qpb}v^*_{\bb p_\qpa}\bigr|^2
    }{\bigl|(\bb p_{\qpa} + \bb p_{\qpb})^2+m_\phi^2-\omega^2\bigr|^{2}}
    ,
\end{equation}
where $\omega_{\bb p_1 +\bb p_2}=(\bb p_1 +\bb p_2)\cdot {\bb v}_\dm - (\bb p_1+\bb p_2)^2/2m_\dm$ is the energy deposited. The quantity
\begin{equation}
    \FBCS(\bb p_1,\bb p_2)
    \equiv \bigl|u^*_{\bb p_1} v_{\bb p_2}+u_{\bb p_2}v^*_{\bb p_1}\bigr|^2
    =
        \frac12\biggl(
            1-\frac{\Erelative_{\bb p_1}\Erelative_{\bb p_2}-\Delta^2}
                   {E_{\qp}(\bb p_1)\,E_{\qp}(\bb p_2)}
        \biggr)
\end{equation}
is the BCS \textit{coherence factor}\footnote{The signs of the various terms in the coherence factor depend on the type of interaction. See, e.g. \cite{Bardeen:1957mv,tinkham:2004}.} \cite{tinkham:2004}. The $\Erelative_{\bb p_1}\Erelative_{\bb p_2}$ term can be dropped if $\bb p_1$ or $\bb p_2$ (or both) are integrated over, provided the remainder of the integrand is even in $\bb p_1$ and $\bb p_2$. But if \textit{e.g.} $\bb p_2$ is fixed to be $\bb q -\bb p_1$, then this term must be kept when integrating over $\bb p_1$.

It is straightforward to see that this matches onto the rate for free electron scattering when $\bb p_1$ and $\bb p_2$ are away from the gap ($|\bb p_i^2/2\mstar-E_\fermi|\gg \Delta$). In this limit, because of the functional form of the coherence factor, we always have $p_\qpa < \sqrt{2{\mstar} E_{\fermi}}$ and $p_\qpb > \sqrt{2{\mstar} E_{\fermi}}$ (or vice versa), with $\FBCS(p_1, p_2)\simeq 1$. Then $E(\bb p_\qpa)\simeq E_\fermi-\bb p_\qpa^2/2\mstar$ and $E(\bb p_\qpb) \simeq \bb p_2^2/2{\mstar} - E_{\fermi}$, so the energy delta function in the rate of \cref{eq:rate} reduces to
\begin{equation}
    \delta\left(\omega- E_\qpa - E_\qpb\right)
    \longrightarrow
    \delta\left[
        \omega - \left(
            \frac{\bb p_\qpb^2}{2{\mstar}} - \frac{\bb p_\qpa^2}{2{\mstar}}
        \right)
    \right]
    .
\end{equation}
That is, the kinematical constraint reduces to that of ordinary non-relativistic scattering. Meanwhile, the quantity 
\begin{equation}
\label{eq:S}
    S(\bb q,\omega) =
    \sum_f|\langle f|\rho_e(\bb q)|0\rangle|^2\delta(\omega-E_f) 
\end{equation}
is known in the literature as the \textit{dynamic structure factor}, in terms of which we have, at lowest order in perturbation theory,
\begin{equation}
    \Im \left(\frac{-1}{\epsilon^{(1)}(\bb q,\omega)}\right) =
    \frac{\pi e^2}{\bb q^2} S(\bb q,\omega)
    \;.
\end{equation}
In our case, the structure factor is simply given by
\begin{equation}
    S(\bb q,\omega) = 
    \int\frac{\du^3\bb p_\qpa}{(2\pi)^3}\frac{\du^3\bb p_\qpb}{(2\pi)^3}\,
    \FBCS(\bb p_\qpa, \bb p_\qpb)
    (2\pi)^3\delta^{(3)}(\bb q - \bb p_\qpa - \bb p_\qpb)
    \delta\Bigl(\omega - E_\qp(\bb p_\qpa) - E_\qp(\bb p_\qpb)\Bigr)
    \;,
\end{equation}
allowing the loss function, $\Im(-1/\epsilon_\BCS)$, to be expressed in terms of the QP dispersion relation and the BCS coherence factor. When higher order terms are included, we can resum the series (at zero temperature) in the \textit{random phase approximation} (RPA)~\cite{nozieres1959electron,mahan2013many}:
\begin{equation}
    \hat{\mathbb{S}}^{(2,\rm RPA)} \supset
    -{g_eg_\dm}\int\!\frac{\du^4q}{(2\pi)^4}\;
    i\frac{\rho_\dm^\dagger(q)\,\rho_e(q)}{q^2-m_\phi^2+i\varepsilon}\,
    \frac{1}{\epsilon\rpa(q)}\;,
\end{equation}
where $\epsilon\rpa(\bb q,\omega) \equiv 1+\chi(\bb q,\omega)$, for $1-\chi(\bb q,\omega)\equiv-1/\epsilon^{(1)}(\bb q,\omega)$. Importantly, we have
\begin{equation}
    \label{eq:rpa}
    \Im \epsilon\rpa(\bb q,\omega) =
    \Im\left(\frac{-1}{\epsilon^{(1)}(\bb q,\omega)}\right) =
    \Im \chi(\bb q,\omega) =
    \frac{\pi e^2}{\bb q^2}S(\bb q, \omega)
    \;.
\end{equation}
Putting everything together, we have
\begin{multline}
    \Gamma\rpa(v_\dm) =
    g^2_eg^2_\dm\!\int\!\frac{\du^3 \bb p_\qpa}{(2\pi)^3}
        \frac{\du^3\bb p_\qpb}{(2\pi)^3}\,
        \frac{
            \bigl|u^*_{\bb p_\qpa} v_{\bb p_\qpb}
                + u_{\bb p_\qpb}v^*_{\bb p_\qpa}\bigr|^2
        }{
            \bigl|(\bb p_{\qpa} + \bb p_{\qpb})^2 + m_\phi^2-\omega^2\bigr|^{2}
        }\,
        \frac{1}{
            |\epsilon\rpa(\bb p_1 +\bb p_2,\omega_{\bb p_1 +\bb p_2})|^2
        }
        \\\times
        2\pi\delta\Bigl(
            \omega_{\bb p_1+\bb p_2}- E_\qp(\bb p_1)\! - E_\qp(\bb p_2)
        \Bigr)
    \;,
\end{multline}
and writing the coherence factor in terms of $\epsilon^{(1)}$ gives 
\begin{align}
    \label{eq:cf-rate}
    \Gamma\rpa(v_\dm) &=
    \int\!\frac{\du^3\bb q}{(2\pi)^3}
        \frac{|V(\bb q)|^2}{|\epsilon\rpa(\bb q,\omega_{\bb q})|^2}
        \frac{2\bb q^2}{e^2} \Im \left(
            \frac{-1}{\epsilon^{(1)}(\bb q,\omega_{\bb q})}
        \right)
    \nonumber\\&=
    \int\!\frac{\du^3\bb q}{(2\pi)^3}
        \frac{|V(\bb q)|^2}{|\epsilon\rpa(\bb q,\omega_{\bb q})|^2}
        \frac{2\bb q^2}{e^2}\Im \left(\epsilon\rpa(\bb q,\omega_{\bb q})\right)
    \nonumber\\&=
    \int\!\frac{\du^3\bb q}{(2\pi)^3}
        |V(\bb q)|^2\frac{2\bb q^2}{e^2}
        \Im \left(\frac{-1}{\epsilon\rpa(\bb q,\omega_{\bb q})}\right)
    \;.
\end{align}
To the accuracy that $\epsilon\rpa$ represents the true dielectric function $\epsilon$, we have derived \cref{eq:overall-rate}.\ignorespaces
\footnote{\ignorespaces
    \begin{minipage}[t]{0.99\textwidth}Note the subtle difference in our approach (following \refcite{nozieres1959electron}) from \refcite{Knapen:2021run}. We define $S(\bb q,\omega)$ to be given strictly by \cref{eq:S}, while $\epsilon\rpa$ has been resummed in perturbation theory. One may optionally redefine $S$ in terms of a resummed density operator $\rho\rpa$ following \refcite{nozieres1959electron}.  \end{minipage}\ignorespaces
} The full form of $\epsilon\rpa$ in the BCS vacuum will be discussed in future work \cite{futureBCS}. In this Letter, we make the approximation
\begin{align}
    \Im \left(\frac{-1}{\epsilon\rpa(\bb q,\omega_{\bb q})}\right)
    \simeq
    \frac{\Im \epsilon\rpa_\BCS(\bb q,\omega_{\bb q})}
        {|\epsilon_{\rm L}(\bb q,\omega_{\bb q})|^2}
    \;,
\end{align}
where the Lindhard function \cite{nozieres1959electron,dressel2002electrodynamics} $\epsilon_\lindhard\equiv \epsilon\rpa_{\rm FEG}$ is the RPA dielectric function for a free electron gas (FEG), and accounts for screening and in-medium effects in a normal metal. Fortunately, by \cref{eq:rpa}, $\Im \epsilon\rpa_\BCS$ can be evaluated from the dynamic structure function without knowing the full form of $\epsilon\rpa_\BCS(\bb q,\omega)$:
\begin{align}
    \Im \left(\epsilon\rpa_\BCS(\bb q,\omega)\right) &=
    \frac{\pi e^2}{\bb q^2} S(\bb q,\omega)
    \nonumber \\&=
    \frac{e^2}{2\bb q^2}
    \int\frac{\du^3\bb p_\qpa}{(2\pi)^3}\frac{\du^3\bb p_\qpb}{(2\pi)^3}\,
    \FBCS(\bb p_\qpa, \bb p_\qpb)
    (2\pi)^4\delta^{(3)}(\bb q - \bb p_\qpa - \bb p_\qpb)
    \delta\Bigl(\omega - E_\qp(\bb p_\qpa) - E_\qp(\bb p_\qpb)\Bigr)
    \;.
\end{align}
This is because the imaginary part depends only on the spectrum of the Hamiltonian and on its relation to the operator $\rho_e$. It corresponds precisely to the sum over states in the scattering rate.

\ssection{Quasiparticle Downconversion}
We simulate QP down-conversion following a similar procedure to the calculations of \refscite{Kaplan:1976zz,kozorezov,Guruswamy_2014,martinis2020saving}. The principal difference is that we track the full momentum vector for scattered particles to retain information about the scattering direction relative to the momentum of the initial scattering event. Here we review the relevant scattering rate calculations, and we derive the emission angles of QPs and phonons in each relaxation process. We compare the final result to the well-established normal metal case described in \refscite{rammer2004quantum,Jacoboni:1983zz}. Down-conversion in the limit of high-energy initial QPs is also discussed by \refscite{kozorezov,OK}.

\begin{figure}\centering
    \includegraphics[width=\textwidth]{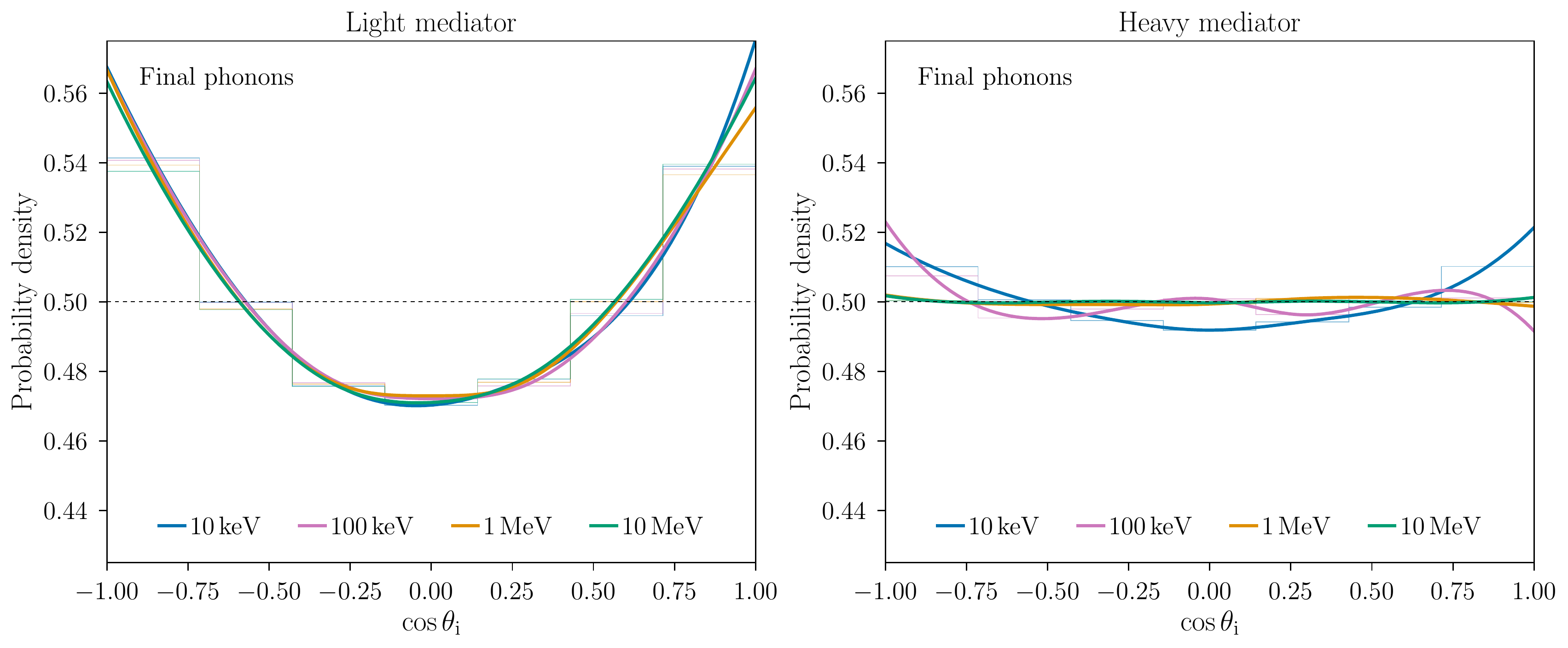}
    \caption{
        Angular distributions of phonons produced by DM scattering in Al. The angles shown are defined with respect to the axis of the DM wind. The distribution of DM orientations in the Standard Halo Model is included. The left and right panel show the distributions in the light- and heavy-mediator limits, respectively. In each panel, the colors correspond to different DM masses, and a dashed horizontal line at $\costh_\qpi = \frac12$ indicates the isotropic distribution. Thick lines interpolate between histogram values (thin lines) for ease of visualization. In the right panel, the curves include only events with total deposit $\omega < 20\Delta$, for which the effects of down-conversion are less significant.
    }
    \label{fig:final-phonons}
\end{figure}

\ssubsection{Phonon Scattering at the Fermi Surface}

We treat QP--phonon interactions following \refcite{Kaplan:1976zz}. The simplified model in this treatment contains a single acoustic phonon branch with the dispersion relation $\omega_{q}= q c_s$, where $\bb q$ is the phonon momentum and $c_s$ is the sound speed in the material. For the dynamics of the problem, scattering is contained to the first Brillouin zone ($q<\frac{2\pi}{a}$), so we do not include an explicit upper limit in momentum in the rate integral. As significant down-conversion already occurs at energies well below the optical phonon modes in most superconductors, we do not explicitly include optical phonon emission in our calculations. This is compatible with the conclusions of past down-conversion codes (see {\it e.g.} \refscite{Guruswamy_2014,kozorezov}).

For acoustic phonon emission, we first adopt the result of \refcite{Kaplan:1976zz} for the emission rate of an acoustic phonon of momentum $q$ in the zero-temperature limit:
\begin{equation} \label{eq:kaplanGamma}
    \frac{\du\Gamma}{\du\omegaph_q}(E_\qp)
    =
        \frac{2\pi}{Z_0}\alpha^2F(q)
        \Re\left(
            \frac{E_\qp-\myhbar\omegaph_q}
                 {\sqrt{(E_\qp-\myhbar\omegaph_q)^2-\Delta^2}}
        \right)
        \times
        \left[1-\frac{\Delta^2}{E_\qp(E_\qp-\myhbar\omegaph_q)}\right]
\end{equation}
where $Z_0$ is the `renormalization parameter' defined in \refcite{Kaplan:1976zz} (Typically $Z_0\sim 2$). We have also assumed that $\alpha^2(q)F(q)\approx b\,\omegaph_q^2$, the Debye solution for this quantity \cite{rammer2004quantum}. The QP energy $E_{\qp}$ is defined in \cref{eq:dispersion}.

\ssubsection{Computing the Scattering Angle}

Given the general scattering rate of \cref{eq:kaplanGamma}, we also need to determine the scattering angle for a phonon of energy $\myhbar\omega_{\bb q}= \myhbar q c_s$ emitted by an QP of energy $E_\qp({\bb k})$. We find this angle by first solving for $k'$ from the conservation of energy relation,
    $E_\qp({\bb k'}) = E_\qp({\bb k})-\myhbar q c_s$,
where $E_{\qp}(\bb k)\equiv E_{\qp}(\myhbar \bb k)$ as defined in \cref{eq:dispersion}. Solving, we get
\begin{equation}
    k'^2 = 2\mstar\left[
        (E_\qp({\bb k})-\myhbar \omegaph_{\bb q})^2 - \Delta^2
    \right]^{1/2} + p_\fermi^2
    .
\end{equation}
In a normal metal or a semiconductor, the $p_\fermi$ dependence cancels. However, for a superconductor with $\Delta > 0$, the $p_\fermi$ dependence is retained.

We can now use momentum conservation to solve for scattering angle. Writing $\bb k = \bb k' + \bb q$, we have $k'^2 = k^2 + q^2 - 2kq\cos\theta_q$. Solving explicitly for $k$ in the QP dispersion relation gives
\begin{equation}
    k = \left[
        2\mstar\left(
            E_\fermi + s\sqrt{E_\qp^2-\Delta^2}
        \right)
    \right]^{1/2}
    ,
\end{equation}
where $s = \pm 1$. An analogous sign $s^\prime$ appears in the solution for $k^\prime$. We then find 
\begin{equation}
    \label{eq:cos-theta-q}
    \cos\theta_q =
    \frac{\hbarsq\omegaph_q^2 + 4\gamma^2\Delta\left[
            s\sqrt{(E^2_\qp - \Delta^2)}
            -s' \sqrt{(E_\qp - \myhbar\omegaph_q)^2 - \Delta^2}
        \right]
    }{
        4\gamma \myhbar\omegaph_q\left[
            E_\fermi\Delta + s \Delta\sqrt{
                (E^2_\qp - \Delta^2)
            }
        \right]^{1/2}
    }
    \ ,\qquad\gamma 
    \equiv\sqrt{\frac{\frac12\mstar c_s^2}{\Delta}}
    \;.
\end{equation}

We now consider a few limiting cases to understand the angular spread in the phonon spectrum. First, observe that for QPs far from the gap, with $\Delta \ll E_\qp \ll E_\fermi$, the scattering angle is unrestricted. In this limit, we have
\begin{equation}
    \cos\theta_q \simeq \frac{1}{4\gamma}
        \frac{\myhbar\omegaph_q}{\Delta}
        \left(\frac{\Delta}{E_\fermi}\right)^{1/2}
        +s' \gamma\left(\frac{\Delta}{E_\fermi}\right)^{1/2}
        + (s - s')\gamma\frac{E_\qp}{\omegaph_q}
            \left(\frac{\Delta}{E_\fermi}\right)^{1/2}
    .
\end{equation}
For typical materials, $\gamma$ is $\mathcal O(1)$ and $E_\fermi \gg \Delta$. Thus, if $s \neq s'$, the last term dominates in the limit of small $\omegaph_q$, and $|\cos\theta_q| = 1$ is allowed. On the other hand, if $s = s'$, then the last term vanishes. The first term can be made arbitrarily small in the small-$\omegaph_q$ limit, and can even be made to cancel with the second term, which is itself always small. In this case, $\cos\theta_q = 0$ is allowed.

On the other hand, consider the near-gap regime, where $k \sim p_\fermi$. Here we can write $E_\qp = (1+\delta)\Delta$ with $\delta \ll 1$, and since the final-state QP energy is at least $\Delta$, we must have $\omegaph_q = a\delta\Delta$ with $0 < a < 1$. Inserting these expressions into \cref{eq:cos-theta-q} and expanding for small $\delta$ gives
\begin{equation}
    \label{eq:cthetaq}
    \cos\theta_q \simeq \frac{
        s\delta - s'\sqrt{1-a}
    }{
        (\delta/2)^{1/2}(a/\gamma)
    }
    \left(\frac{\Delta}{E_\fermi}\right)^{1/2}
    .
\end{equation}
However, the phonon emission process is kinematically forbidden if $\left|\cos\theta_q\right| > 1$, and minimizing \cref{eq:cthetaq} over $a$ gives
\begin{equation}
    \min_{a\in(0, 1)}\cos\theta_q \simeq
    \gamma\left(\frac{E_\fermi}{\Delta}\right)^{-3/2}
    \left(s\frac{E_\fermi}{\Delta}\sqrt{\frac{2}{\delta}} - 1\right)
    .
\end{equation}
Thus, phonon emission is only kinematically allowed for sufficiently large $\delta$, i.e., for
\begin{equation}
    \label{eq:delta-min}
    \delta \gtrsim \delta_{\mathrm{min}} \equiv
    \frac{2\gamma^2(E_\fermi/\Delta)^2}
    {\left[(E_\fermi/\Delta)^{3/2} + \gamma\right]^2}
    .
\end{equation}
This result is self-consistent: in typical materials, $\gamma$ is $\mathcal O(1)$, and $E_\fermi \gg \Delta$, so $\delta_{\mathrm{min}} \ll 1$. This gives rise to a condition for phonon emission:
\begin{equation}
    \label{eq:qp-ballisticity}
    E_\qp \gtrsim \Delta +
    \frac{4c_s^2E_\fermi^2\mstar}
        {\left(2E_\fermi^{3/2} + \Delta\sqrt{2\mstar c_s^2}\right)^2}\Delta
    .
\end{equation}
QPs with energies below this threshold are ballistic: no phonon emission is allowed.

The angular distribution of final-state phonons is peaked oppositely to that of final-state QPs, and is closer to the isotropic distribution. The distribution is shown explicitly in \cref{fig:final-phonons} for the same cases as in \cref{fig:angular-distribution-summary}.

\ssubsection{Relation to Normal Metal Scattering}

It is instructive to compare \cref{eq:kaplanGamma} to the equivalent rate in the normal metal case. In a normal metal, this emission rate becomes \cite{rammer2004quantum}
\begin{equation}\label{eq:scatteringRate}
    \Gamma(E_{\bb k}) =
    2\pi\int \frac{\du^3\bb k'}{(2\pi\myhbar)^3}
    \left|\alpha_{\bb q}\right|^2
    \delta(E_{\bb k} - E_{\bb k'} - \myhbar\omega_{\bb q})
    \delta^{(3)}(\bb k-\bb k'-\bb q)
    ,
\end{equation}
where $\alpha_{\bb q}$ is the coupling for electron--phonon scattering. If we assume that scattering is isotropic, we can make a change of variables such that $\int\du\bb p' \propto \int\du \omegaph_q \dd(\cos\theta)$, where $\cos\theta$ is the scattering angle between $\bb k$ and $\bb k'$. For scattering near the Fermi energy in the metal, conservation of momentum gives
\begin{equation}
    \label{eq:momentum}
    q^2 = k^2 + k'^2 - 2kk'\cos\theta
    \approx 2p_\fermi^2(1 - \cos\theta)
\end{equation}
where $p_\fermi = \sqrt{2mE_\fermi}$ is the Fermi momentum. This implies $\du(\cos\theta) = -q\dd q/p_\fermi^2$, which in turn allows us to write the differential scattering rate for $\myhbar\omegaph_q < E_\qp$ as
\begin{equation}
    \frac{\du\Gamma_n}{\du \omegaph_q}(E_\qp) = 2\pi\alpha^2F(q)\;
    .
\end{equation}
Here the subscript `$n$' indicates the normal metal case; $N_0$ is the normal metal density of states at $E_\fermi$; and $\alpha^2F(q)$ is the coupling-weighted phonon density of states, given by
\begin{equation}
    \alpha^2F(q) = \frac{N_0}{2\myhbar p_\fermi^2}
    \int_0^{q_{\mathrm{max}}}\du q'\, q' |\alpha_{q'}^2|\delta(q'-q)
    \;.
\end{equation}
 For the scaling used earlier this gives the normal result that the acoustic scattering rate goes as $E_\qp^3$ when integrated over energy, and the differential spectrum goes as $\omegaph_q^2$.

When we derive the emission rate for superconductors, there are two important modifications required to get from \cref{eq:scatteringRate} to the final differential rate in emitted phonon energy. First, we modify the dispersion relation to that of the QPs in the superconductor. In the metal, we had $E_{\bb k} = \Erelative_{\bb k} \equiv \hbarsq\bb p^2/(2\mstar)-E_{\fermi}$ relative to the Fermi surface, but in a superconductor, the gap energy modifies this to
\begin{equation}
    \label{eq:dispSC}
    E_\qp({\bb k}) = \sqrt{\Erelative_{\bb k}+\Delta^2}
    \;.
\end{equation}
In the normal metal phonon interaction, the total coupling has a density of states term that is valid in the metal, but not in the superconductor, since there are no states at $E_{\bb k} < \Delta$. We can use the fact that the total number of states is the same to find the modified density of states at a given energy, i.e., we have $\du E\, N_s(E) = \du\Erelative\, N_n(\Erelative)$, where the subscript `$s$' indicates the superconductor case. We thus find that
\begin{equation}
    N_s(E) \approx N_0\frac{\du\Erelative}{\du E}
    = N_0\frac{E}{\sqrt{E^2-\Delta^2}}
    .
\end{equation}
Thus, to rescale $\alpha^2F(q)$ for the superconducting case, we have to rescale by the ratio of superconducting to normal states at a given energy. This gives us the first additional factor in the superconducting rate equation. \refcite{tinkham:2004} points out that, in principle, the gap function is complex-valued, hence the need to take only the real part.

The second correction factor is the coherence factor described in the previous section. This is a purely BCS effect that arises from the collective nature of the superconducting states. This factor ensures that the divergence in the density of states does not lead to a divergence in the phonon scattering rate. Taking the type I coherence factor for phonon emission from \refcite{tinkham:2004}, we find
\begin{equation}
    \FBCS(\Delta,E_\qp,\omegaph_q) \approx
    \frac{1}{Z_0}\left(
        1 - \frac{\Delta^2}{E_\qp(E_\qp-\myhbar\omegaph_q)}
    \right)
    ,
\end{equation}
using the same normalization procedure as \refcite{Kaplan:1976zz}. Combining this with the previous correction factor and multiplying by the normal metal scattering rate produces the scattering rate of \cref{eq:kaplanGamma}. This heuristic argument elucidates the origin of \cref{eq:kaplanGamma} in the low-temperature limit with a less formal approach than that in \refcite{Kaplan:1976zz}.

\ssection{Estimation of directional reach}
In this section, we detail the methodology used to produce \cref{fig:reach} in the main text, and we demonstrate the impact of total-deposit cuts on the directionality of the signal.

\ssubsection{Statistical methods}
The directional reach in \cref{fig:reach} is estimated in two distinct ways. The dashed curves are based on measurement of an asymmetry in the counts of final-state QPs between two bins of equal solid angle, and the dotted curve is based on comparison of the full angular distribution against the null hypothesis.

\subsubsection{Two-bin reach}
We first discuss the two-bin reach estimate. The premise of this test is that an isotropic background gives rise to an isotropic distribution of final-state quasiparticles, and, in particular, produces statistically-indistinguishable counts in any two bins of equal solid angle. Since the DM wind is not isotropic, it is possible to statistically distinguish the counts in the two bins given a sufficient number of events.

We make the simplifying assumption that the angles of the final-state QPs are independent random variables. This is not strictly the case, since QPs that originate from the same event have some angular correlation. However, given a large number of events, such correlations are extremely sparse. Moreover, by simulating an ensemble of \textit{isotropic} DM scattering events, we have directly checked that such correlations are irrelevant at the number of events needed to establish directionality. Having made this assumption, the assignment of an angular bin to each QP can be treated as a Bernoulli trial. We can then use the binomial test to determine whether to reject the isotropic distribution given a particular set of QPs.

This procedure allows us to determine whether a particular sample of final-state QPs is consistent with an isotropic signal. Next, we must translate this to a minimum number of events needed to establish directionality. To that end, we randomly draw samples of $N_\qp = 2,3,4,\dotsc$ QPs and evaluate the binomial test for each sample, repeating the process many times for each fixed $N$ to obtain a median $p$-value. We advance $N_\qp$ until this median $p$-value drops below the threshold value of $0.05$, and we interpret the resulting value of $N$ as the typical number of QPs needed in order to establish directionality. Finally, this number of QPs must be translated to a number of scattering events. We estimate this number as $N_{\mathrm e} \equiv N_\qp / \bar n$, where $\bar n$ is the average number of final-state QPs produced by a scattering event. This $N_{\mathrm{e}}$ is indicated by the dashed curves in \cref{fig:reach}.

For the heavy mediator case, we introduce an additional step. As we discuss below, the distribution of final-state QPs produced by an event approaches the isotropic distribution as the deposit becomes much larger than the superconducting gap $2\Delta$. Thus, it is advantageous to restrict attention to events with total deposit below some cut, even at the cost of a reduced event rate below the cut. The dashed curve in the right panel of \cref{fig:reach} is a composite of two reach curves obtained with cuts $\omega < 10\Delta$ and $\omega < 50\Delta$. Due to the complicated relationship between the total deposit, the initial directionality, and the effects of down-conversion, each of these cuts preserves overall directionality for a different mass range, and the combination of the two gives directional sensitivity over the entire mass range. Due to the large deposits favored in the heavy-mediator case, a cut on the total deposit is essential to establish directionality for all but the lowest masses.

This simplistic treatment produces a rough upper bound on the number of events needed to detect directionality, and admits a very direct interpretation. Even at the level of a two-bin experimental configuration, more sophisticated statistical treatments may yield slightly stronger results. In particular, it is possible to extract directionality using the Skellam distribution, as in rate modulation experiments \cite{Blanco:2021hlm}. Here one treats the count in each bin as a Poisson random variable, so that the difference in the number of counts between the two bins is a Skellam-distributed random variable. It is then possible to test whether the two Poisson counts are produced with the same rate. However, our case is simpler than a traditional rate modulation experiment in that the target is isotropic, so the total rate is fixed. We have checked that using the Skellam distribution offers a slight enhancement to the two-bin reach, but does not qualitatively change the result.

\subsubsection{Full angular distribution}
\Cref{fig:reach} also includes an estimate based on the full angular distribution of the final-state QPs, assuming an experimental configuration with great angular precision. For this estimate, we begin with an ensemble of simulated DM scattering events. Next, a second `null' ensemble of DM scattering events is simulated with an isotropic distribution of DM directions, using the same speed distribution as the Standard Halo Model. For each simulated event, we compute the mean angle of the final-state QPs, $\langle\costh_\qpi\rangle$. We thus obtain two sets of samples $\{\langle\costh_\qpi\rangle\}_{\mathrm{SHM}}$ and $\{\langle\costh_\qpi\rangle\}_{\mathrm{null}}$ for the two ensembles. We then determine the average number of events needed to reject at 95\%~C.L. the hypothesis that the `SHM' and `null' samples are drawn from the same distribution, using the two-sample Kolmogorov--Smirnov test. Note that unlike the binomial test of the previous case, the samples being compared in this case are truly independent: since each mean angle corresponds to a single event, and vice versa, all of the values we draw originate from different scattering events, and they are thus independent random variates.

For light mediators, the two procedures give a nearly identical result, and \cref{fig:reach} shows only the simpler two-bin result. This is easily understood in light of the typical number of QPs produced in a scattering event: for a light mediator, a typical final state consists of an $\mathcal O(1)$ number of QPs, and deviation of these QPs from the isotropic distribution generally gives rise to a two-bin asymmetry. In fact, for large DM masses, taking the mean angle in each event discards information to the point that the distributional test slightly \textit{underperforms} the two-bin test. For a heavy mediator, on the other hand, a typical final state may consist of $\mathcal O(\num{e4})$ QPs for large DM masses. The angular mean $\langle\costh_\qpi\rangle$ will typically be very close to zero, and distinguishing the distribution of these means from the isotropic case becomes a problem of precision measurement. The large number of QPs makes this a realistic possibility. However, we estimate that achieving the high-resolution dotted curve in \cref{fig:reach} would require a measurement precision of $\mathcal O(\num{e-2})$ in $\costh_\qpi$.

\ssubsection{Energy dependence of directionality}

\begin{figure*}\centering
    \includegraphics[width=\textwidth]{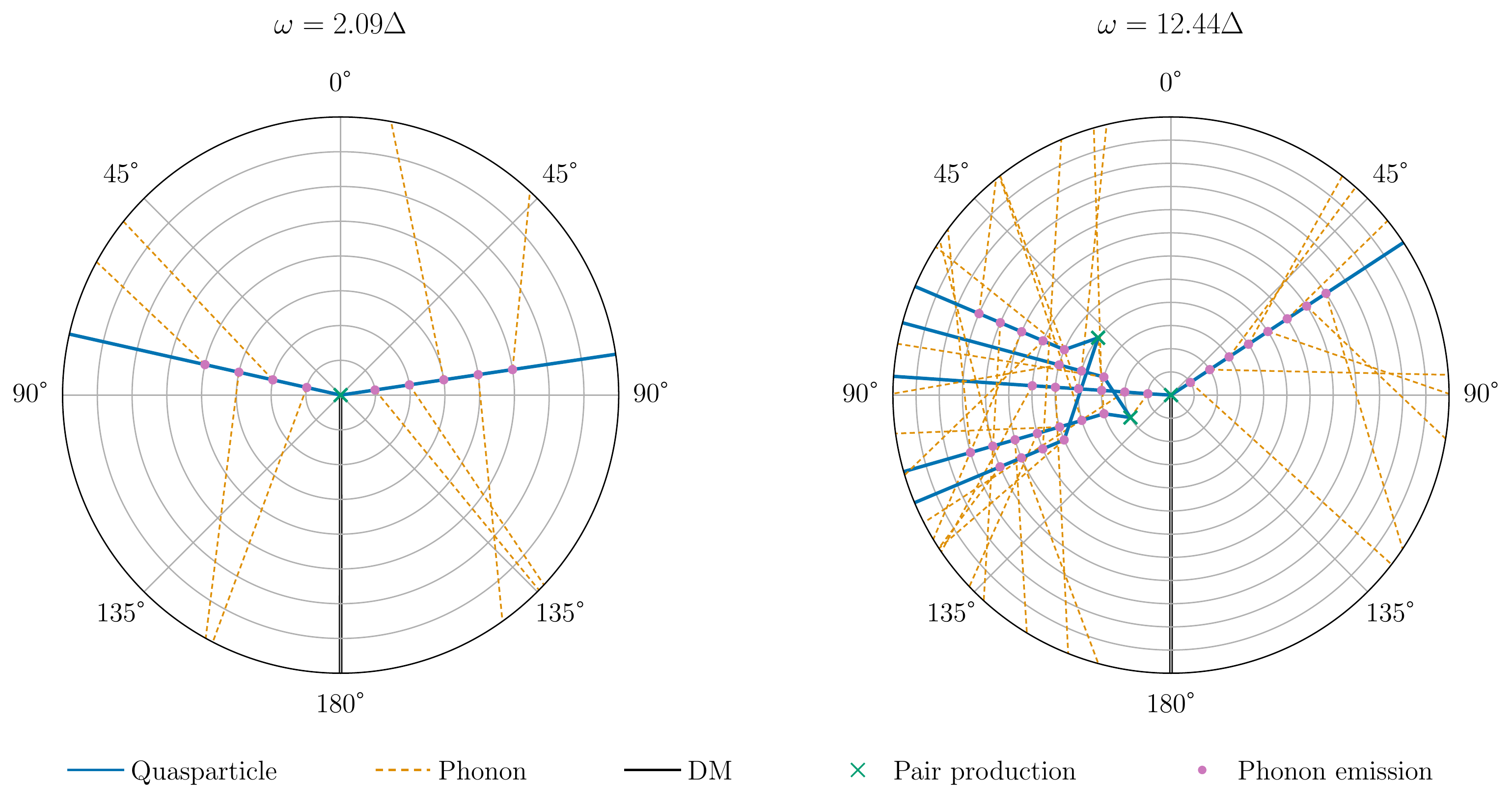}
    \caption{Simulation of the QP down-conversion process for two initial energy deposits. The black line indicates the direction of the incoming DM. For visualization purposes, each line terminates at an angular coordinate corresponding to its angle from the incoming DM axis. Thus, the directions of the plotted excitations are represented by their endpoints, not by their slopes. \textit{Left:} since $\omega = 2.09\Delta$, neither initial QP can emit a phonon with energy above $2\Delta$, so no phonons can produce QP pairs. Thus, the only QPs in the final state are those produced in the DM scattering event, with their directions barely altered. \textit{Right:} now $\omega = 12.44\Delta$ and the emission of above-gap phonons is allowed. Thus, a chain of phonon emissions and decays erases much of the initial directionality, although a preference for off-axis final states remains visible.}
    \label{fig:down-conversion-scattering}
\end{figure*}

In this section, we demonstrate the dependence of final-state directionality on the deposited energy. As noted in the main text, larger deposits allow for a larger number of relaxation events during the down-conversion process, which attenuates the correlation between the directions of the final-state excitations and those of the initial excitations produced by DM scattering. This is illustrated for particular realizations in \cref{fig:down-conversion-scattering}. In the left panel, the relatively small number of relaxation events and the low energies of the emitted phonons guarantee that the directions of the initial QPs are well-preserved, and no additional QPs are produced. In the right panel, on the other hand, the larger deposit allows for a larger number of relaxation events, with additional QP pair production. While there remains a directional correlation between the final state and the initial QPs, this correlation is partially erased by down-conversion. Directional information is effectively lost for very large deposits.

The impact of down-conversion means that small deposits are favorable for directionality even if the directionality of the initial QPs is smaller in this regime. We can demonstrate this explicitly by evaluating the final-state asymmetry as a function of the deposited energy. To facilitate quantitative discussion of directionality, we introduce a quantitative `two-bin asymmetry' $\mathcal A_2$, defined as follows. As discussed above, we divide the final-state QPs into two bins of equal solid angle: the `on-axis' bin, with $\left|\costh\right| > \frac12$, and the `off-axis' bin, with $\left|\costh\right| < \frac12$. We denote the counts in these two bins by $n_{\mathrm{on}}$ and $n_{\mathrm{off}}$, respectively, and then define
\begin{equation}
    \mathcal A_2 \equiv
        \left|
        2\times\frac{n_{\mathrm{on}}}{n_{\mathrm{on}} + n_{\mathrm{off}}} - 1
        \right|
    .
\end{equation}
In particular, the isotropic distribution gives $\mathcal A_2 = 0$, and a totally asymmetric distribution gives $\mathcal A_2 = 1$.

\begin{figure}\centering
    \includegraphics[width=\textwidth]{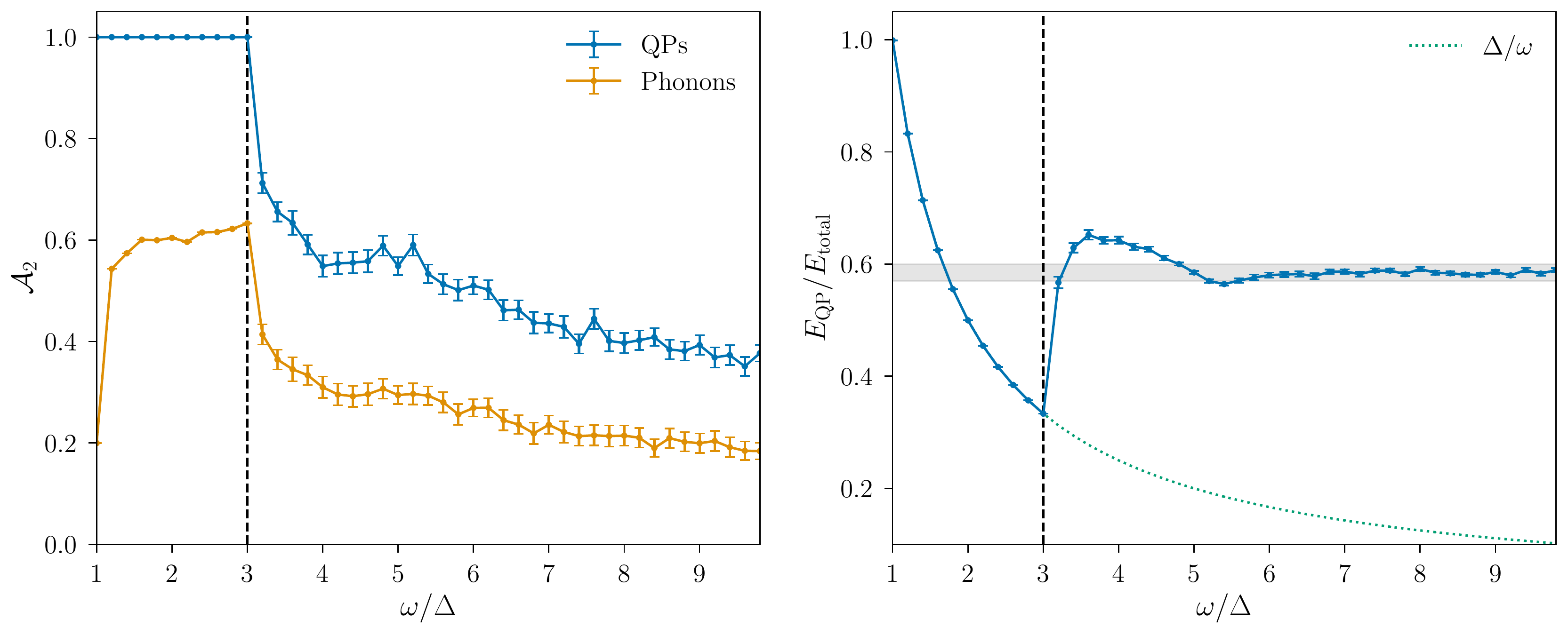}
    \caption{Directionality of final-state QPs resulting from the down-conversion of a single QP of energy $\omega$ oriented with $\costh = 1$. \textbf{Left:} asymmetry $\mathcal A_2$ of final-state QPs (blue) and phonons (orange). Directionality of the final excitations is quickly erased for $\omega \gg \Delta$, and $\mathcal A_2$ approaches zero in the large-$\omega$ limit. \textbf{Right:} Fraction of the total energy residing in the QP system after down-conversion. The gray band shows the range of asymptotic results 0.57--0.60 obtained in the literature for bulk Al superconductors \cite{Guruswamy_2014,kozorezov}.}
    \label{fig:energy-asymmetry}
\end{figure}

The left panel of \cref{fig:energy-asymmetry} shows $\mathcal A_2$ for the final-state QPs produced by a single QP injected at fixed energy $\omega$ with $\costh = 1$. This serves as a proxy for the preservation of directionality at fixed deposit. Directionality is almost perfectly preserved for very low-energy QPs with $\omega < 3\Delta$. Above this threshold, it becomes possible for the QP to emit an above-gap phonon, with $E_\ph > 2\Delta$. Such a phonon subsequently decays to another pair of QPs, which have only weak angular correlation with the original QP. The nature of the $3\Delta$ threshold is also clearly visible in the right panel of \cref{fig:energy-asymmetry}, which shows the fraction of the total energy that resides in the QP system after down-conversion. For $E_\qp < 3\Delta$, the QP relaxes almost all the way to the gap by emission of sub-gap phonons, which cannot produce any additional QPs. Thus, the final state consists of a single QP with $E_\qp \approx \Delta$, and a set of phonons with all the remaining energy from the deposit. The fraction of the initial energy in the QP system is approximately $\Delta / \omega$. Upon reaching $\omega > 3\Delta$, emission of above-gap phonons produces additional QPs in the final state, sharply raising the fraction of the total energy in the QP system. At large $\omega$, this fraction asymptotically reaches $\sim\!0.6$. This is consistent with previous studies of down-conversion in the high-energy limit, which find fractions between $0.57$ and $0.60$ \cite{Guruswamy_2014,kozorezov} (gray band in \cref{fig:energy-asymmetry}).

\begin{figure}\centering
    \includegraphics[width=\textwidth]{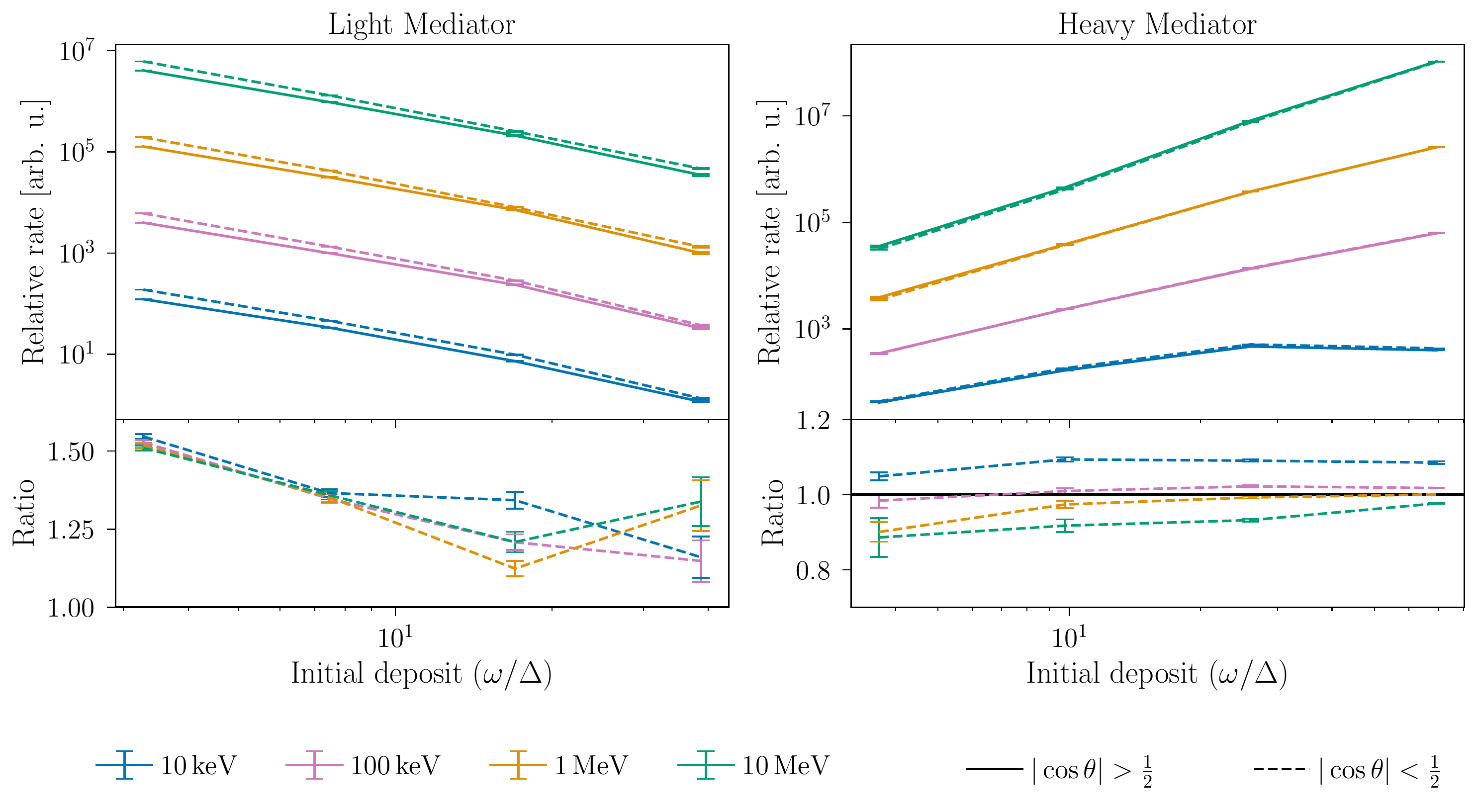}
    \caption{\textit{Top panels}: total event rates as a function of initial deposit in the on-axis ($\left|\cos\theta\right|>\frac12$, solid) and off-axis ($\left|\cos\theta\right|>\frac12$, dashed) bins for several DM masses, for light (\textit{left panel}) and heavy (\textit{right panel}) mediators. Normalization of each pair of curves is arbitrary and fixed for ease of visualization. \textit{Bottom panels:} ratio of off-axis to on-axis QP counts. The left and right panels assume a light and heavy mediator, respectively.}
    \label{fig:relative-rates}
\end{figure}

Naively, \cref{fig:energy-asymmetry} suggests that the smallest deposits will yield the strongest directionality. However, imposing an upper limit on the deposit also influences the directionality of the initial excitations, and, of course, the rate of events which fall below the cut. To study the total directionality as a function of the deposited energy, in \cref{fig:relative-rates}, we show the spectrum of final-state QPs in each of the two angular bins (`on-axis' and `off-axis') for several DM masses in the light- and heavy-mediator limits. The bottom panels of \cref{fig:relative-rates} show the ratios of these spectra, i.e., a signed and shifted version of the two-bin asymmetry $\mathcal A_2$.

For light mediators, directionality is quickly lost for deposits well above the gap, and the ratio approaches 1. Moreover, the ratio is generally above 1. For heavy mediators, due to the directionality of the initial excitations, the ratio declines less noticeably, but it is near 1 throughout the plot and asymptotically reaches 1. As anticipated in \cref{fig:angular-distribution-summary}, light mediators always enhance the off-axis rate, while heavy mediators can enhance either the off-axis or on-axis rates, depending on the DM mass: since scattering through a heavy mediator can produce a QP distribution peaked either in the forward direction or off-axis, the ratio can be either below or above 1.

\end{document}